%% file: ir_model.tex
\documentclass[useAMS,usenatbib,usegraphicx]{mn2e}


 \newcommand{\gal}{\sc morgana}
\newcommand{\gs}{\sc grasil} \newcommand{\be}{\begin{equation}}
  \newcommand{\ee}{\end{equation}} \newcommand{\bea}{\begin{eqnarray}}
  \newcommand{\eea}{\end{eqnarray}}

\newcommand{\lir}{$L_{\rm IR}$}
\newcommand{\lirgs}{$L^{\rm GS}_{\rm IR}$}

\def\lesssim{\,\lower2truept\hbox{${<\atop\hbox{\raise4truept\hbox{$\sim$}}}$}\,}
\def\gtrsim{\,\lower2truept\hbox{${>\atop\hbox{\raise4truept\hbox{$\sim$}}}$}\,}

\title[Dust modeling in SAMs II] {Evaluating and Improving
  Semi-analytic modeling of Dust in Galaxies based on Radiative
  Transfer Calculations II: Dust Emission in the Infrared.}

\author[Fontanot et al.]{
\parbox[t]{\textwidth}{ 
Fabio Fontanot$^{1,2}$ \& Rachel S. Somerville$^{3,4}$
}
\vspace*{6pt}\\
  $^1$INAF-Osservatorio Astronomico, Via Tiepolo 11, I-34131 Trieste, Italy \\
  $^2$MPIA Max-Planck-Institute f{\"u}r Astronomie, K{\"o}nigstuhl 17, 69117 Heidelberg, Germany\\
  $^3$Space Telescope Science Institute, 3700 San Martin Drive, Baltimore, MD 21218, USA \\
$^4$Johns Hopkins University, Baltimore, MD 21218, USA \\
  email: fontanot@oats.inaf.it}

\begin{document}
\date{Accepted ... Received ...}

\maketitle
\begin{abstract} 
  Interstellar dust grains are responsible for modifying the spectral
  energy distribution (SED) of galaxies, both absorbing starlight at
  UV and optical wavelengths and converting this energy into thermal
  emission in the infrared. The detailed description of these
  phenomena is of fundamental importance in order to compare the
  predictions of theoretical models of galaxy formation and evolution
  with the most recent observations in the infrared region. In this
  paper we compare the results of {\sc grasil}, a code explicitly
  solving for the equation of radiative transfer in a dusty medium,
  with the predictions of a variety of IR template libraries, both
  analytically and observationally determined. We employ star
  formation history samples extracted from the semi-analytical galaxy
  formation model {\gal} to create libraries of synthetic SEDs from
  the near- to the far-infrared. We consider model predictions at
  different redshift ranges to explore any possible influence in the
  shape and normalisation of the SEDs due to the expected evolution of
  the galaxy properties. We compute the total absorbed starlight
  predicted by {\gs} at optical wavelengths to statistically compare
  the synthetic SEDs with the selected IR templates. We show that
  synthetic SEDs at a given total infrared luminosity are predicted to
  be systematically different at different redshift and for different
  properties of the underlying model galaxy. However, we determine
  spectral regions where the agreement between the results of
  radiative transfer and IR templates is good in a statistical sense
  (i.e. in terms of the luminosity functions). Moreover, we highlight
  some potentially relevant discrepancies between the different
  approaches, both in the region dominated by PAH emission and at
  sub-mm wavelengths. These results determine potentially critical
  issues in the infrared luminosity functions as predicted by
  semi-analytical models coupled with different IR flux estimators.
\end{abstract}


\section{Introduction}
\label{sec:introduction}
The great advance in our understanding of the spectral energy
distributions (SEDs) of galaxies has clearly demonstrated the
ubiquitous presence of dust in galaxies. Dust plays a major role in
galaxy evolution: it contributes to setting the physical conditions
for star formation in galaxies, both seeding the formation of
molecules and shielding the molecular gas from ionising radiation. It
is also important in the processes of cooling and condensation of giant
molecular clouds (MCs) to the densities required for the onset of star
formation \citep[see][for a review]{DorschnerHenning95}. At the same
time, the presence of dust has a strong impact on the intrinsic
spectral energy distributions (SED) of galaxies, since dust grains
efficiently absorb and scatter radiation at short wavelengths
($\lambda<1\mu m$). The absorbed energy is then thermally re-emitted
at longer wavelengths in the infra-red (IR) region ($\lambda>1\mu m$).

Dust-related processes of attenuation and re-emission are key
mechanisms in understanding and interpreting galactic SEDs. Estimates
based on the results of the IRAS satellite demonstrated that the IR
region is responsible for at least $\sim 30\%$ of the bolometric
luminosity of nearby galaxies \citep{Popescu02}.  Early IR surveys
\citep[see, e.g.][]{Sanders96} discovered a population of heavily
obscured luminous and ultra-luminous infrared galaxies (LIRGs, with IR
luminosities \lir $\sim 10^{11}-10^{12} L_\odot$, and ULIRGs,
\lir $> 10^{12}$). Moreover, the study of the cosmic infrared
background \citep[see][for a review]{HauserDwek01} shows that the
global energy emitted in the IR is comparable to the direct starlight
emission, detectable mainly at optical wavelengths, emerging from
galaxies.  Moreover, galaxy number counts in the mid-IR are an order
of magnitude higher than those predicted by non-evolving local
luminosity functions \citep{Pozzetti98, Elbaz99, Lagache99}. Several
groups \citep{Chary01, Dole01, Lagache03} have shown that all these
pieces of evidence favour strong evolution of the bright IR
sources. This has been confirmed with direct observations of
high-redshift sources using facilities like ISO \citep{Dole01,
  Elbaz02, Gruppioni02}, the Spitzer Space Telescope \citep{LeFloch05,
  Babbedge06} and SCUBA \citep{Chapman03, Chapman05}. This strong
evolution of luminous IR sources can also account for the peak in the
cosmic infrared background.

Dust reprocessed radiation therefore represents a fundamental aspect
of galactic emission: in particular a significant fraction of star
formation activity is expected to be heavily extinguished and
detectable only in the IR, since the molecular clouds are both the
sites for star formation and among the most dusty environments. Most
interestingly, this contribution was probably more important at higher
redshift, where star formation activity and dust reprocessing are both
stronger. The detailed modeling of dust properties is then a key issue
in order to understand galaxy evolution at different cosmic
epochs. Our understanding of these processes is complicated by the
strong dependence of galactic SEDs on the relative geometry between
dust and stars \citep{Granato00, Tuffs04}. In fact, the youngest
stars, which dominate the UV luminosity, are expected to be heavily
extinguished by their optically thick parent Molecular Clouds
(MC). Older stars survive the disruption of the MCs and in fact, an
age-dependent extinction has been postulated in the cirrus component
for the intermediate-age stars \citep{Popescu00, Panuzzo07}, in order
to reproduce the observed properties of disc-dominated galaxies. The
emerging SED depends critically on the temperature distribution in the
different environments (cirrus and MCs), as different dust species
(with different composition and size) respond differently to the
radiation field.

Therefore, a detailed treatment of the effects of dust grains in
different spectral regions (both attenuation in the optical and
re-emission in the IR) is of fundamental importance in order to infer
the physical properties of galaxies from multi-wavelength observations
and, conversely, to simulate galaxy UV-to-IR fluxes and colours from
the predictions of theoretical models, both semi-analytic and
numerical simulations. Semi-analytic models \citep[SAMs, e.g.,
][]{Bower06, Croton06, DeLucia06, Monaco07, Somerville08} simulate the
formation and evolution of galaxies, moving from a statistical
description of the evolution of the Dark Matter structure (``merger
trees'' either analytically determined or extracted from the
predictions of numerical simulations) and using a well defined set of
simple, but physically motivated, analytic ``recipes'' to describe the
evolution and interplay of the various processes acting on the
baryonic component (i.e. gas cooling and infall, star formation,
stellar and AGN feedback). SAMs provide detailed information on the
physical properties of model galaxies (in terms of star formation and
AGN activity, stellar mass and gas mass content, stellar and gas
metallicity) and some information on the relative geometry of the
system (the sizes of the disc and bulge components). Stellar
population synthesis codes \citep[e.g.,][]{FiocRV97, Silva98,
  Leitherer99, Bruzual03}, coupled with the star formation histories
predicted by the SAM, may be used to produce synthetic spectra and
photometry for the unattenuated starlight. However, most SAMs do not
follow the evolution of the relationship between dust mass and gas and
metal content, nor the relative geometry between stars and dust, nor
the detailed dust properties (grain size distribution and
composition).  In the past years, several groups (see e.g
\citealt{Devriendt00,Hatton03,Baugh05,Silva05,Fontanot07b,Lacey08})
have presented SAM predictions with refined treatments for dust
attenuation and emission based on different assumptions for dust grain
properties and the relative star/gas/dust geometry.

The various spectrophotometric codes presented in the literature show
reasonable agreement on the intrinsic (pure-stellar) galactic SEDs,
but they differ in the treatment of the highly complex effects of
dust. Assuming that all energy absorbed by the dust at UV and optical
wavelengths is re-radiated in the IR, it is possible to predict the
expected IR fluxes in the region of interest. The approaches that have
been used to date can be divided into two broad categories: those that
use codes explicitly solving the equations of radiative transfer (RT)
in a dusty medium, and those that make use of analytically motivated
and/or empirically calibrated IR template libraries. Clearly the two
approaches differ in the complexity and (therefore) in the
computational effort involved in the treatment of dust effects. IR
templates are usually built up by combining the results of analytic
calculations \citep{Desert90,Dale02,Lagache04} and/or available
observational constraints \citep{Chary01,Rieke09}. They are then
calibrated by means of comparison with a well defined set of low-z
objects. It is therefore unclear to which extent the same templates
provide a good description of the SED properties at higher redshifts:
several authors \citep[see e.g][]{Rieke09} discuss possible
limitations in using these templates outside the redshift range on
which they are calibrated. RT-solver methods of course provide more
detailed predictions and a more correct dependence on the physical
properties and geometrical configurations of dust in
galaxies. However, the use of a full RT-solver coupled to a SAM
substantially reduces the efficiency and flexibility of the
semi-analytic approach (becoming the bottleneck of the computation,
see e.g. \citealt{Fontanot07b}). Furthermore, this approach requires,
and is sensitive to, a large number of additional parameters
(specifying the details of the dust model) which must still be chosen
somewhat empirically and which carry along numerous assumptions (such
as constancy with redshift and/or environment). Given these drawbacks,
and bearing in mind the large number of uncertainties and
approximations already inherent in the SAM modeling, it is worth
asking whether the difference between the results obtained with a full
RT treatment versus IR templates justifies the use of the former tool.

This paper is the second in a series aimed at better understanding and
characterising the different approaches for determining dusty SEDs in
the SAM framework. In a first paper \citep[][hereafter Paper
  I]{Fontanot09a} we studied the effect of different prescriptions for
dust attenuation at ultra-violet and optical wavelengths. In this
paper we will instead focus on the re-emission of the absorbed energy
in the infrared region (from $2 \mu m$ to the sub-mm). In the light of
the previous discussion, we set out here to address a number of
issues. First we briefly recap the study that we carried out in Paper
I, in order to understand how well the amount of light absorbed by
dust is predicted by analytic recipes for dust attenuation compared
with a RT-solver. Our main goal in then to compare, on a statistical
basis, the SED shapes contained in various IR template libraries from
the literature with the predictions of a full RT solver coupled with a
SAM and dust model. Both aspects of SED modeling will be of
fundamental importance, given the growing interest triggered by
infrared observations and the future availability of large datasets,
thanks to facilities like the Herschel satellite. As a final goal for
this project, we identify critical spectral regions, where the choice
between RT-solver and IR templates may bias the comparison of model
predictions with available observational data.  To achieve these
goals, we make use of the {\gs} RT code coupled with the {\gal}
semi-analytic model. We derive a set of IR SED templates from the
coupled {\gal}+{\gs} outputs, binned as a function of bolometric
luminosity, and compare these with a range of templates from the
literature. Finally, we compare the IR luminosity functions predicted
by the full {\gal}+{\gs} runs with those that we obtain if we
substitute the template approach. It is worth stressing that {\gs} is
of course not the only RT-solver available in the literature. Several
other groups have developed alternative approaches and techniques (see
e.g \citealt{Popescu00,Dopita05,Jonsson06}) for the prediction of
panchromatic SEDs, based on different assumptions for the properties
of star forming regions, dust composition and distribution as well as
for the star/gas/dust geometry. A detailed comparison of the
performances of these codes, applied to the same sample of star
formation histories, would be extremely interesting but is clearly
beyond the scope of the present paper. Nonetheless, in the following
we will refer to the {\gal}+{\gs} predictions as being representative
for the performance of a generic SAM+RT solver code, and leave further
comparison between the different approaches to future work.

This paper is organised as follows. In sec.~\ref{sec:models} we
describe the main features of the {\gal} and {\gs} models and we
review the procedure used in Paper I to define the synthetic RT-based
SED libraries that we will use in the comparison. We introduce the IR
template libraries in sec.~\ref{sec:templates} and we summarise the
main results of Paper I in sec.~\ref{sec:f08}. In
sec.~\ref{sec:comp_seds} we present the statistical comparison between
the properties of the RT-based SEDs and the IR templates, while in
sec.~\ref{sec:irlf} we discuss the effect of the different
prescriptions on the statistical properties of model galaxies through
the analysis of IR luminosity functions. Finally in
section~\ref{sec:final} we present our conclusions.

\section{Galaxy models}

\subsection{Semi-analytic model: {\gal}}

\label{sec:models}
\begin{figure*}
  \centerline{
    \includegraphics[width=9cm]{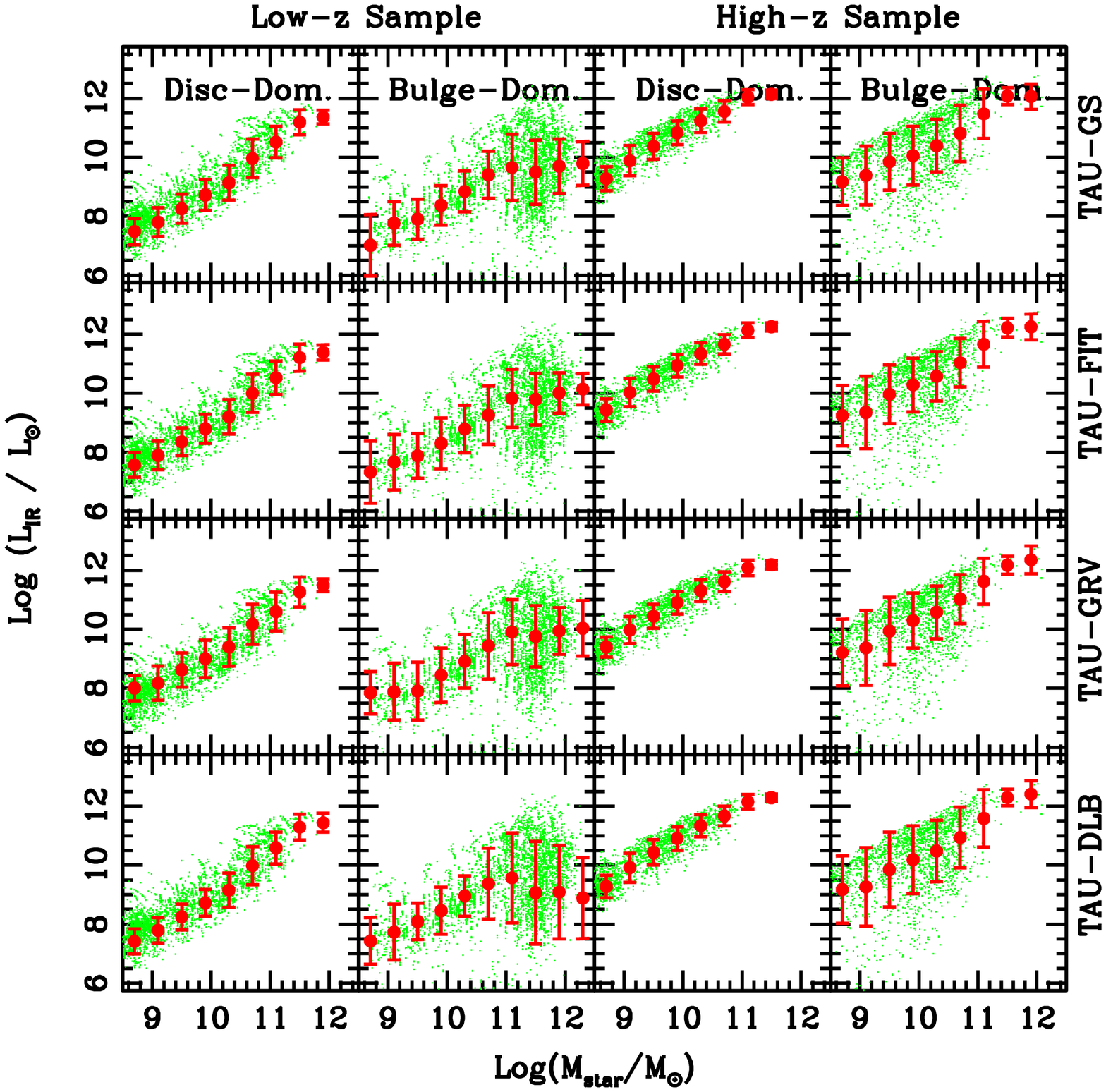}
    \includegraphics[width=9cm]{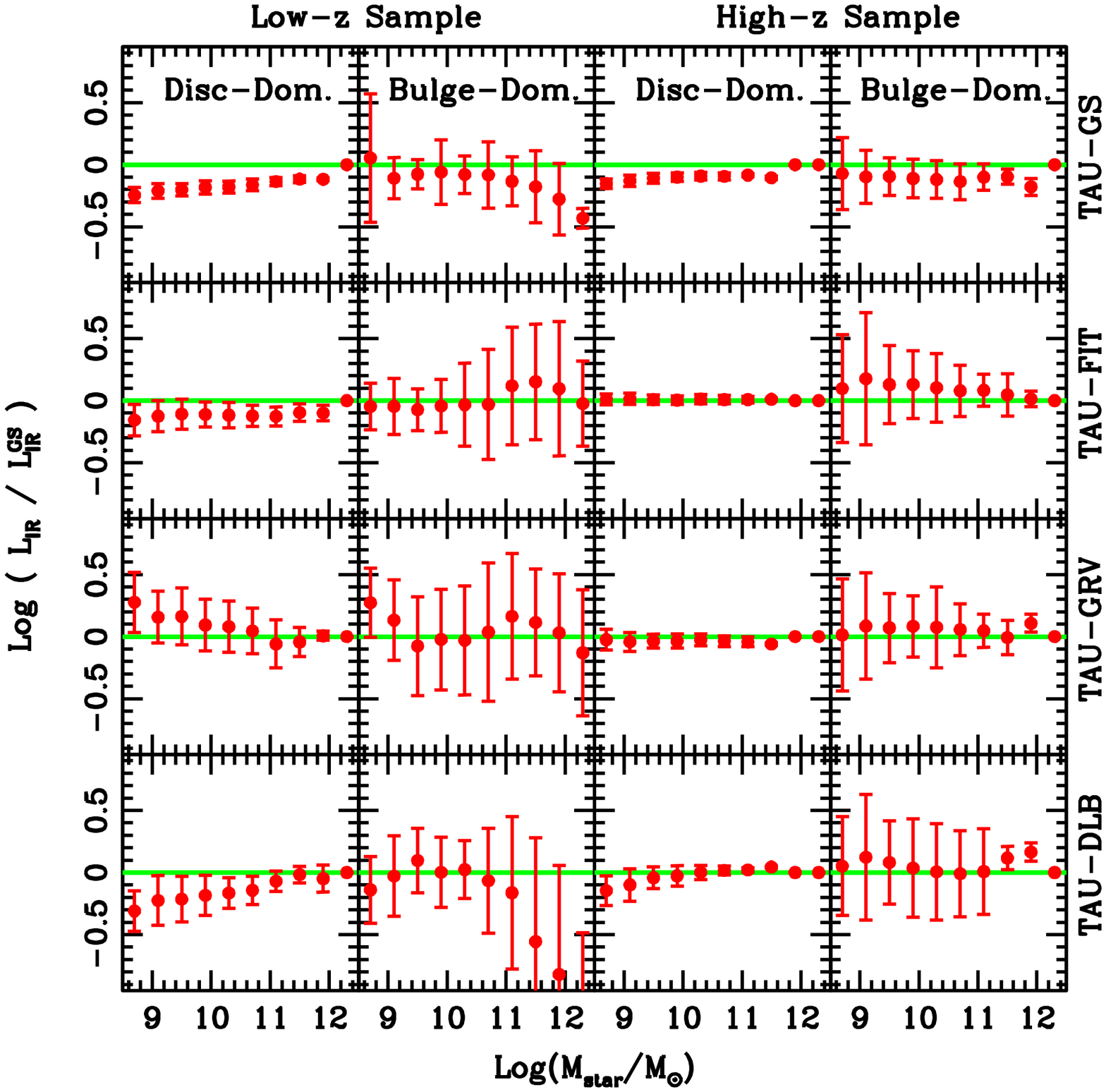}
  }
  \caption{Left panels: absorbed starlight \lir as a function of
    stellar mass for the objects in our samples. Green dots refer to
    the {\gs} results, while red circles with errorbars represent the
    mean predictions of the 4 prescriptions presented in Paper I, as
    indicated in the labels. Right panels: residuals in the relations
    (symbols and labels as in the left panels).}
    \label{fig:figL}
\end{figure*}

We make use of the semi-analytic model {\gal}, developed by
\citep{Monaco07} and \citet{Fontanot07b}.  We refer the reader to
these papers and Paper I for more details on the galaxy formation
model, and here we just recall its main features: the model implements
a sophisticated treatment of mass and energy flows between the
different gas phases (cold, hot and stars) and galactic components
(bulge, disc and halo), as well as a new treatment for gas cooling and
infall (following \citealt{Viola08}).  It also includes both a
multi-phase description of star formation and feedback (following
\citealt{Monaco04}) and a self consistent description of AGN activity
and feedback \citep{Fontanot06}.  For consistency with Paper I, we use
the same star formation histories (SFHs) and SED samples\footnote{In
  Paper I we refer to these ensembles as {\it ML libraries}. For the
  sake of simplicity and to avoid further confusion with the
  definition of the IR template libraries, in the following we will
  simply refer to the SFHs we extract from {\gal} as {\it samples}
  (i.e. low-z and high-z samples).}, extracted from the {\gal}
\citep{Monaco07} realization presented in \citet{Fontanot07b}.  In
this study we are mainly interested in the IR emission due to the
coupling of stellar activity with the dusty interstellar medium, and
we do not explicitly include the contribution of the central AGN to
the predicted SEDs. This particular {\gal} realization assumes a
\citet{Salpeter55} Initial Mass Function with mass range from 0.1 to
100 $M_\odot$.

Every model galaxy is represented assuming a composite geometry
including both a spheroid and a disc component. Disc exponential
profiles are computed following the \citet{MoMaoWhite98} formalism:
the spin parameter of the DM halo is randomly extracted from a well
defined distribution and the angular momentum is conserved. Bulge
sizes are computed assuming that the kinetic energy is conserved in
merger events \citep{Cole00}. The presence of a bulge is taken into
account when disc sizes are computed. {\gal} provides predictions for
the star formation history, metal enrichment, and mass assembly of
each component separately. This information is then interfaced with
the RT-solver {\gs} \citep{Silva98} in order to predict the resulting
SED from the UV to the Radio. The {\gal} realization we consider is
able to reproduce the local and $z=1$ stellar mass function, the
cosmic star formation history, the evolution of the stellar mass
density, the slope and normalisation of the Tully-Fisher relation for
spiral discs, the redshift distribution and luminosity function
evolution for $K$-band selected samples, and, more interestingly, the
number counts of $850$ micron selected sources. Despite these
successes, the agreement between model predictions and observations of
the apparent ``downsized'' trend of galaxy formation is still under
debate (see \citealt{Fontanot07b, Fontanot09b} for a complete
discussion about ``downsizing'' trends and SAMs). It is well
established that this model overpredicts the number of faint, low-mass
galaxies at $z<2$ \citep{Fontana06,Fontanot09b} and the space density
of bright galaxies at $z<1$ \citep{Monaco06}. Moreover, it does not
reproduce the observed levels of star formation activity as a function
of stellar and halo mass in the local universe \citep{Kimm08}.

Many relevant properties of galaxies show redshift evolution,
including star formation rates, sizes, metal and dust content. In
order to test the effects of the predicted changes in these physical
quantities, we follow the same approach as in Paper I, i.e. we draw
two different samples of SFHs from the {\gal} predictions: a low-z
($0.0<z<0.2$) and a high-z sample ($2<z<3$). In Paper I, model
galaxies in each sample are also split into disc-dominated and
bulge-dominated subsamples\footnote{according to their bulge-to-total
  ratio, with a threshold value of 0.6} to test the effect of the
composite geometry. We check that the shape of the IR SED does not
show any strong dependence on the details of the geometrical
configuration and our results are not affected by the
splitting. Unless otherwise explicitly stated, in the following we
will lift this subdivision.

\subsection{RT solver: {\gs}}
\label{sec:grasil}

For each object in our samples we interface {\gal} predictions with
the spectro-photometric code {\gs} (\citealt{Silva98}, for subsequent
improvement we refer the reader to \citealt{Silva99, Granato00,
  Bressan02, Panuzzo03, Vega05}), to compute the corresponding SEDs
from the UV to the Radio. {\gs} solves the equation of RT, taking into
account a state-of-the-art treatment of dust effects. Stars and dust
are distributed in a bulge (King profile) + disc (radial and vertical
exponential profiles) axisymmetric geometry. The clumping of both
(young) stars and dust through a two-phase interstellar medium with
dense giant star-forming molecular clouds embedded in a diffuse
(``cirrus'') phase are considered. The stars are assumed to be born
within the optically thick MCs and to gradually escape from them on a
time-scale $t_{\rm esc}$, giving rise to age- (wavelength-) dependent
extinction (i.e the youngest and most luminous stars suffer larger
extinction than older ones). The dust composition consists of graphite
and silicate grains (with a distribution of grain sizes), and
Polycyclic Aromatic Hydrocarbons (PAH) molecules. At each point within
the galaxy and for each grain type the appropriate temperature $T$ is
computed (either the equilibrium $T$ for big grains or a probability
distribution for small grains and PAHs\footnote{The detailed PAH
  emission spectrum has been updated in \citet{Vega05} based on the
  \citet{LiDraine01} model.}). The radiative transfer of starlight
through dust is computed along the required line of sight yielding the
emerging SED. The simple stellar population (SSP) library
\citep{Bressan98,Bressan02} includes the effect of the dusty envelopes
around AGB stars, and the radio emission from synchrotron radiation
and from ionised gas in HII regions. Although {\gs} computes the
resulting SEDs along different lines of sight, in the following we use
only angle-averaged SEDs.

Most of the physical information (such as stellar mass, star formation
history, cold gas mass, gas metallicity) and parameters (scale radii
for stars and dust in the disc and bulge components) needed by {\gs}
are provided directly by {\gal}. We fixed the parameters used by {\gs}
and not provided by {\gal} as in \citet{Fontanot07b} and Paper I.  (i)
The disc scale heights for stars and dust are set to $0.1$ times the
corresponding scale radii. (ii) We set the escape time-scale of young
stars for the parent MCs to $t_{\rm esc} = 10^7$ yr. This is a good
compromise between the value needed to explain the SED of spirals
($\sim$ a few Myr) and starbursts \citep[$\sim$ a few 10 Myr,
  see]{Silva98} and it is similar to the estimated destruction time
scale of MCs by massive stars. In \citet{Fontanot07b} this value was
found to produce good agreement with the K-band luminosity functions
and the $850 \, \mu$m counts. (iii) The total gas mass is split
between the dense and diffuse phases, assuming that $50\%$ of the gas
is in the star forming molecular clouds ($f_{\rm MC}=0.5$). The
results are not very sensitive to this choice.  (iv) The mass of dust
is obtained by the gas mass and the dust-to-gas mass ratio
$\delta_{\rm dust}$ which is set to evolve linearly with the
metallicity given by the galaxy model, $\delta_{\rm dust}=0.45 \; Z$.
(v) The optical depth of MCs depends on the ratio $\tau_{MC} \propto
\delta_{\rm dust} \, M_{\rm MC}/r_{\rm MC}^2$; we set the mass and
radius of MCs to typical values for the Milky Way, $M_{\rm MC} = 10^6
M_\odot$ and $r_{\rm MC} = 16$ pc. (iv) The dust grain size
distribution and composition is chosen to match the mean Milky Way
extinction curve.

Many authors have suggested that dust properties may evolve as a
result of the evolution of galactic properties and/or differential
metal enrichment (see e.g. \citealt{Schurer09}). In a recent paper
\citet{LoFaro09} analysed the luminosity function of Lyman Break
galaxies at $z>3$ in the {\gal} framework, assuming the same {\gs}
choices for the dust properties and their relation with gas and metal
content. Their results show that the predicted extinctions are larger
than the estimates of \citet{Bouwens07}. They then suggest that
different values for $t_{\rm esc}$ and $f_{\rm MC}$ are able to
reproduce the correct amounts of attenuation. Given the great
uncertainties in the expected properties of dust as a function of
redshift, we assume the same relations between dust and gas content,
as well as the same dust composition and grain distribution for both
the high-z and low-z sample. This assumption is similar in spirit to
applying an IR template library derived from low-redshift observations
at higher redshift. Thus, any differences that we find in the IR SEDs
as a function of redshift are due to the changing physical properties
of the model galaxies, not to changes in the dust properties. Given
the strong dependence of the predicted SEDs on the details of the SFHs
of model galaxies, the quantitative results we present in this paper
are strictly valid only for the {\gal}+{\gs} algorithm, i.e. by
interfacing {\gs} with another SAM, we expect to obtain a different
sample of synthetic SED libraries. Nonetheless, the comparison between
the predictions of these particular SED libraries and IR templates
will provide us fundamental insight on the dependence of SAM
predictions on the different IR modeling. Moreover, as shown in
\citet{Fontanot09b}, the global evolution of bulk galaxy properties
(such as stellar masses and star formation rates) in {\gs} agrees well
with other SAMs in the literature, suggesting that the results
presented here would likely be similar for any currently available
SAM.

\subsection{IR template libraries}
\label{sec:templates}

\begin{figure*}
  \centerline{ 
    \includegraphics[width=15cm]{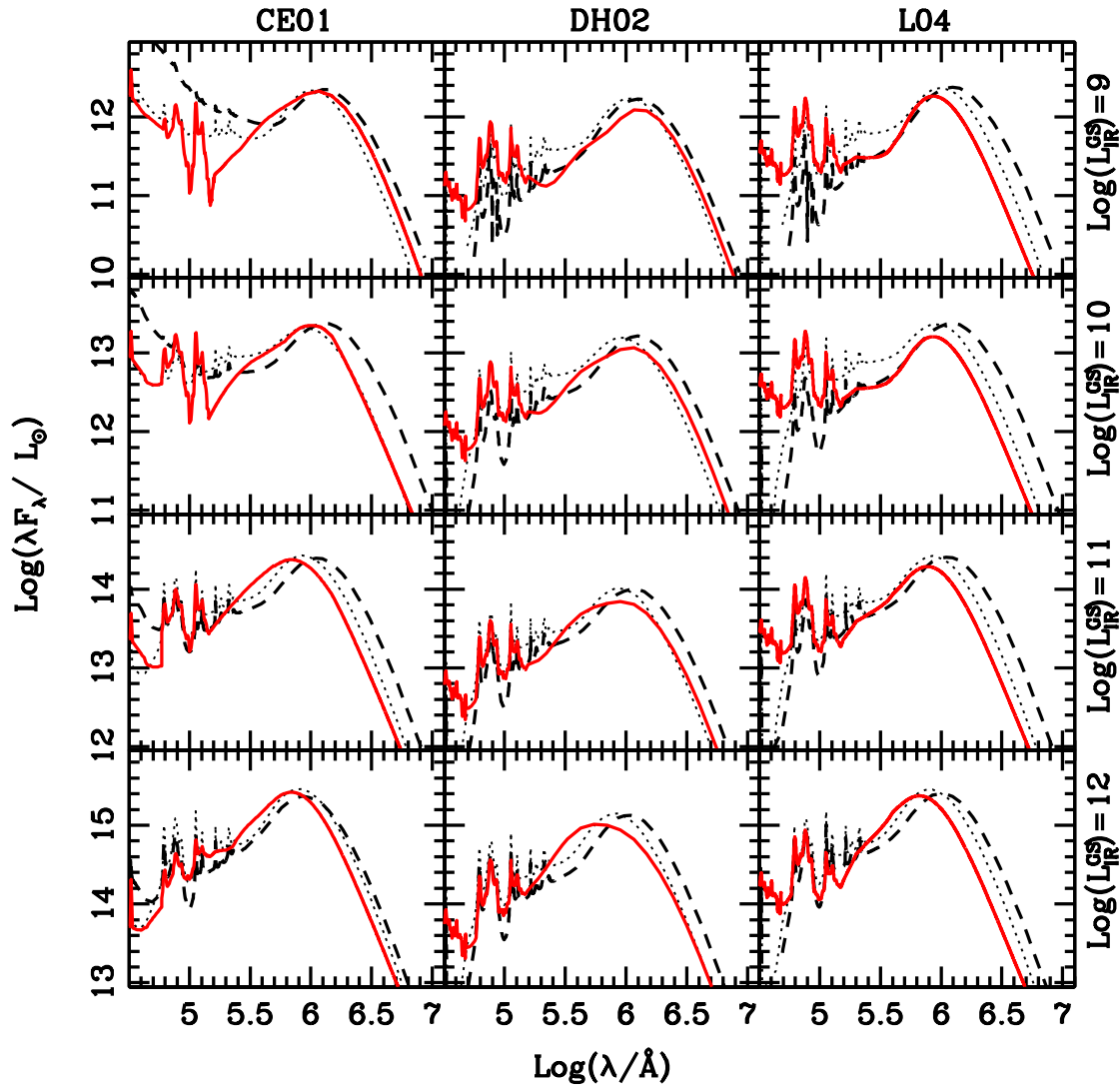} 
  }
  \caption{IR templates from the literature (red solid lines) compared
    to the {\gs} predicted mean SEDs for the CE01, DH02 and L04
    libraries and four representative values of \lir. In each
    panel the dashed and dotted lines refer to the low-z and high-z
    sample respectively. It is worth noting that the apparently
    different shapes of the mean SEDs in the various panels are due to
    the different \lir binning considered in the original
    libraries (see text for more details).}
  \label{fig:sed_mean}
\end{figure*}
\begin{figure*}
  \centerline{ 
    \includegraphics[width=15cm]{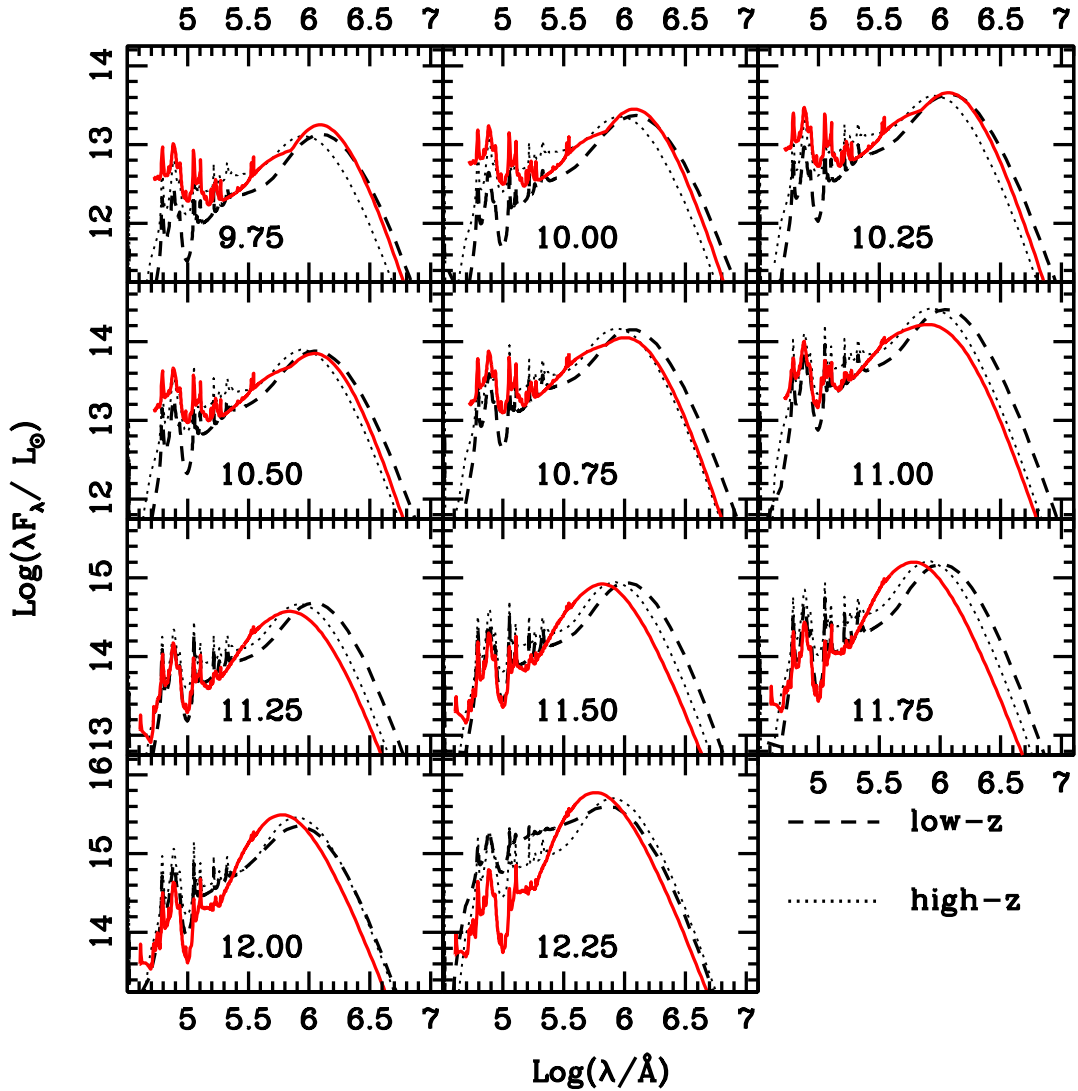} 
  }
  \caption{R09 IR templates (red solid lines) compared with the {\gs}
    predicted mean SEDs. In each panel the \lir reference interval
    is indicated, while dashed and dotted lines refer to the low-z and
    high-z sample respectively.}
  \label{fig:r09_sed_mean}
\end{figure*}
\begin{figure*}
  \centerline{ 
    \includegraphics[width=9cm]{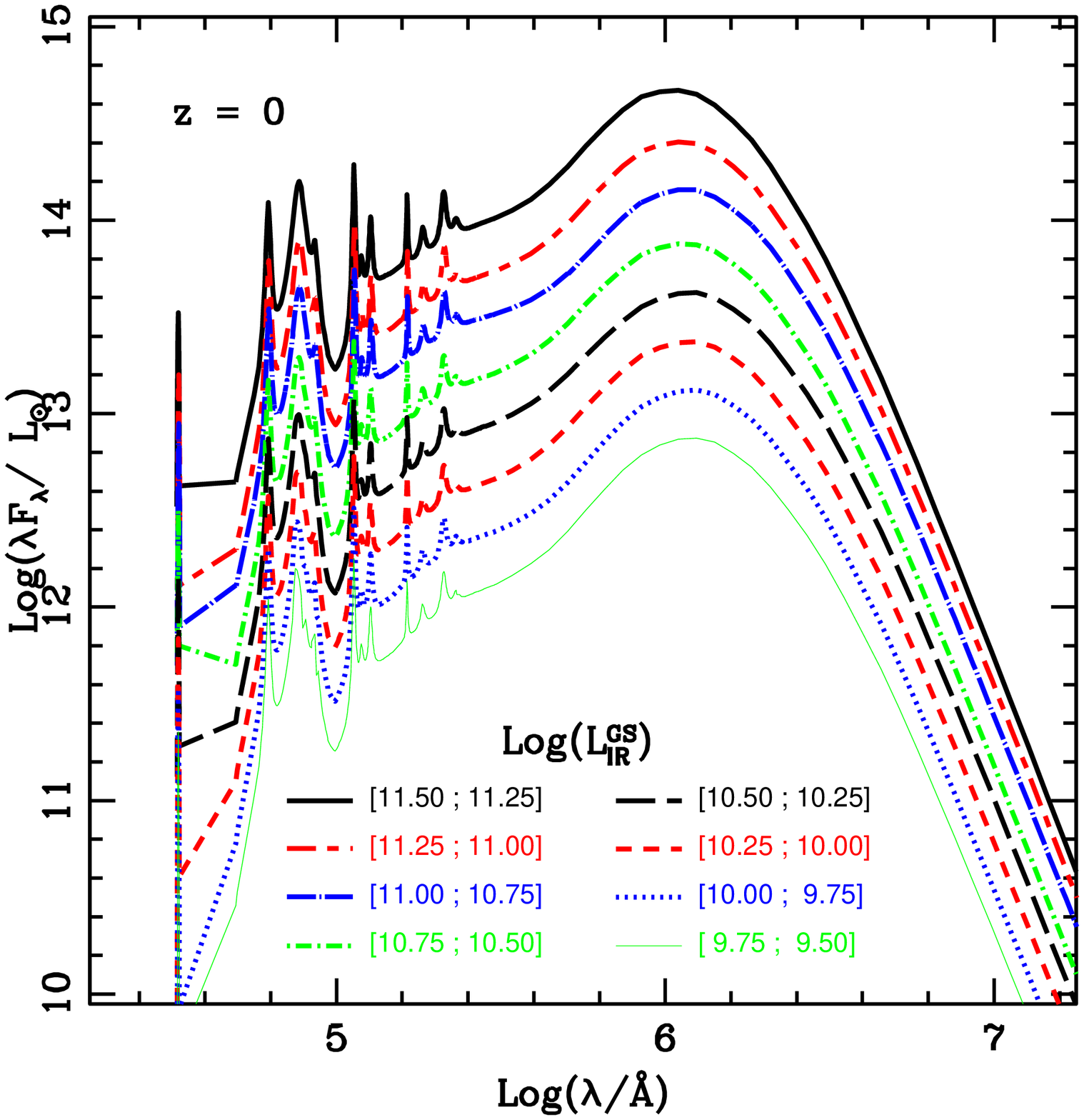}
    \includegraphics[width=9cm]{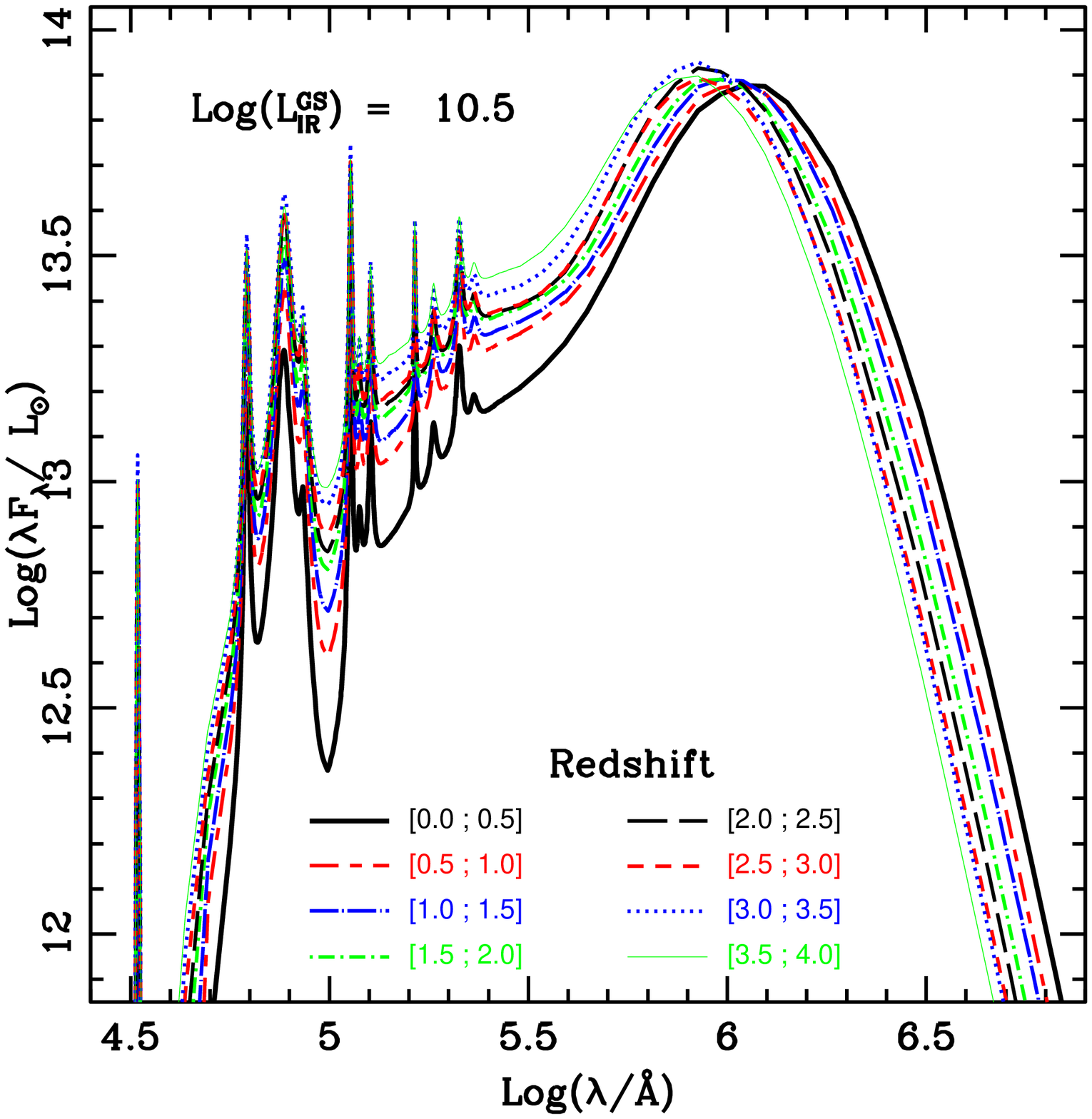}
  }
  \caption{{\it Left Panel:} Mean SED evolution as a function of
    \lir in the low-z sample. {\it Right Panel:} The evolution of
    {\gs} predicted mean SEDs as a function of redshift.}
  \label{fig:lir_evo}
\end{figure*}

Starting from the pioneering work by \citet[hereafter
  DSP90]{Desert90}, many authors proposed analytic solutions to RT
equations, and used their results to define template IR SEDs as a
function of \lir, calibrated by comparison with known local
prototypes. The original formulation of DSP90 was meant to reproduce
the dust emission from individual regions within the Milky-Way. They
assumed 3 main contributors to dust emission at IR wavelengths:
polycyclic aromatic hydrocarbons (PAHs), very small grains and big
grains. The former two are composed of graphite and silicates, with
small and big grains probably dominated by graphite and silicate
respectively. The thermal properties of each species are defined by
its size distribution and its thermal state. Big grains are assumed to
be in near thermal equilibrium. Their emission can be modeled as a
modified black-body spectrum. On the contrary, small grains and PAHs
are very likely in a state that is intermediate between thermal
equilibrium and single photon heating. They are therefore subject to
temperature fluctuations and their emission spectra are much broader
than a modified black-body spectrum.

In the original DSP90 paper, this approach was successfully applied to
reproduce the SED of the Milky Way. \citet[hereafter
  DGS99]{Devriendt99} expanded this framework to define IR template
libraries: since the detailed size distributions are modeled using
free parameters, it is possible to calibrate them by requiring the
model to fit a series of observational constraints, such as the
extinction/attenuation curves, observed IR colours and the IR spectra
of local galaxies. Once the free parameters are fixed it is possible
to use this model to predict the IR spectral contribution
corresponding to each species, embedded in radiation fields of
different intensity. The distribution of dust mass over heating
intensity is usually assumed to follow a power-law. The main advantage
of this approach lies in the relatively small number of parameters
required by the model (with respect to a RT solver). On the other
hand, it relates a single SED to a class of model galaxies (defined by
their infrared luminosity \lir), irrespective of their morphologies or
other properties. Alternative implementations have been proposed by
\citet[hereafter DH02]{Dale02} and \citet[hereafter L04]{Lagache04}.
These implementations differ from the original DSP90 work, mainly in
the modeling of the radiation field (DGS99, DH02), and in the relative
contribution and shape of the IR emission of the different species
(L04, DH02). The DGS99 template library consists of nine SEDs,
corresponding to \lirgs values spanning the range $10^6 L_\odot<$ \lir
$<10^{14} L_\odot$; L04 consists of five SEDs in the range $10^9
L_\odot<$ \lir $<10^{13} L_\odot$; DH02 contains sixty-four SEDs in
the range $2 \times 10^8 L_\odot<$ \lir $<2 \times 10^{14} L_\odot$.

An alternative approach has been proposed by \citet[hereafter
  CE01]{Chary01}. They start from four reference SEDs generated using
{\gs} and constrained to reproduce the observed SED of Arp220,
NGC6090, M82 and M51 from the far-UV to the sub-mm \citep{Silva98}.
These four galaxies are considered representative of four galactic
populations or luminosity classes: ULIRGs, LIRGs, starburst and normal
galaxies respectively. They calibrate the $3 \mu m < \lambda < 18 \mu
m$ region of the four model spectra with ISOCAM observations, then
they split each SED into a mid-IR and a far-IR section and interpolate
between the reference spectra to generate a set of sample template
SEDs at intermediate luminosities.  Additional templates from
\citet{Dale01} are used to widen the range of spectral shapes. The
final templates are then chosen by requiring the two sections to
account for a variety of observational constraints and merging
together the best fit solutions in the two regions. The CE01 library
consists of 105 templates, covering the range $2.7 \times 10^8
L_\odot<$ \lir $<3.5 \times 10^{13} L_\odot$.

Finally, we also consider the recent work of \citet[hereafter
  R09]{Rieke09}. They assemble detailed SEDs for eleven LIRGs and
ULIRGs from observational datasets (including published ISO, IRAS and
NICMOS data as well as previously unpublished IRAC, MIPS and IRS
observations) and use them to build representative average IR
templates. In regions where no homogeneous spectral coverage is
available, they use {\sc galaxev} \citep{Bruzual03} synthetic spectra,
calibrated by requiring them to fit the available photometry. To model
the far infrared SEDs, they assume a single black body with
wavelength-dependent emissivity, while in the Radio they use a single
power law wavelength dependence, as suggested by observed data. The
R09 library includes fourteen SEDs covering the $5.6 \times 10^9
L_\odot<$ \lir $< 10^{13} L_\odot$ range.

\subsection{Analytic Models of Dust Attenuation}
\label{sec:f08}

In Paper I, we considered the effect of dust at optical to
near-infrared wavelengths as predicted by {\gs}, by comparing the
synthetic SED to the intrinsic starlight emission of each model
galaxy. We compute mean dust attenuations and optical-to-near-infrared
colours in the two samples and we then compare them to the predictions
of simpler analytic prescriptions, commonly used in the SAM framework.
These prescriptions are based on (i) an analytic attenuation law,
normalised to (ii) an assumed value for the face-on optical depth at a
reference wavelength (usually in the $V$-band, $\tau_V$), eventually
computed as a function of the physical properties of the galaxy
itself; (iii) an inclination correction, performed by means of simple
analytic models representing simplified geometries (such as the slab
model or the oblate ellipsoid). In Paper I we tested different choices
and combinations of these three elements. In particular, we
considered: (i) the Milky Way \citep{Mathis83}, the \citet{Calzetti00}
and the age dependent \citet{CharlotFall00} attenuation laws, plus the
composite\footnote{i.e the Milky Way extinction curve describes the
  effect of the diffuse ``cirrus'' dust and the \citet{CharlotFall00}
  power-law the attenuation law experienced by young stars in the
  dense birth clouds.} attenuation law of \citet{DeLucia07b}; (ii) the
fitting formulae for $\tau_V$ proposed by \citet{GRV87} and
\citet{DeLucia07b}; (iii) slab and oblate ellipsoid geometries. Our
results show that {\gs} RT calculations can be reproduced with
reasonable agreement even by a model as simple as a slab geometry
combined with the composite age-dependent attenuation curve suggested
by \citep{DeLucia07b}, if an accurate estimate of the intrinsic
$\tau_V$ is used to normalise the attenuation law. Using these
synthetic SED samples, in Paper I we tested different analytic
expressions, relating $\tau_V$ with the physical properties of the
model galaxy (such as bolometric luminosity, stellar mass,
metallicity, cold gas fraction and/or surface density). We
demonstrated that the intrinsic value for $\tau_V$, as predicted by
{\gs}, correlates strongly with the gas metallicity, the cold gas mass
and the scale radius of the model galaxies and we provided simple
analytic fitting formulae.  We then compared our formulae with
analogous prescriptions found in the literature, and in particular
with the the \citet{GRV87} and \citet{DeLucia07b} fitting formulae, by
defining four different analytic prescriptions for dust attenuation
and testing them against {\gs} predictions. The four prescriptions
share the same choice for attenuation law and geometry (i.e. a
\citet{DeLucia07b} composite attenuation law combined with a slab
model), but assume a different normalisation (i.e. $\tau_V$). A first
prescription uses the intrinsic $\tau_V$ value as computed from the
comparison of attenuated and unextinguished {\gs} synthetic SEDs
(TAU-GS). In the remaining we assume both the \citet{GRV87} and
\citet{DeLucia07b} $\tau_V$ fitting formulae plus our results: we then
define TAU-GRV, TAU-DLB and TAU-FIT prescriptions respectively. We
showed that the agreement among the different prescriptions is
satisfactory for the low-z sample, but the discrepancies are
significant in the high-z sample. Our proposed fitting formulae
provide the best compromise in reproducing the {\gs} RT-predictions in
both samples at the same time.

\section{Results}
\label{sec:results}

\subsection{Total Absorbed Starlight}
\label{sec:abs_star}

In this paper we extend the analysis of Paper I by considering the
energy re-emitted by dust at longer wavelength. Therefore, as a first
critical test we compare the total amount of absorbed starlight from
the UV to the near-infrared (\lir) both in {\gs} and in the four
analytic extinction prescriptions. As described in the previous
section, in defining our prescriptions we assume a fixed combination
for the geometry and the attenuation law. It is worth stressing that
alternative analytic dust models presented in the literature do not
necessary assume fixed attenuation laws: for example in \citet{GRV87}
a dependence of the shape of the attenuation law on metallicity is
explicitly considered. We compute the intrinsic value of \lirgs by
comparing the extinguished $F_\lambda^{abs}$ and unextinguished
$F_\lambda^{unabs}$ SEDs:

\be L^{\rm GS}_{\rm IR} = \int_{100 nm}^{2200 nm} (F_\lambda^{unabs} -
F_\lambda^{abs}) d \lambda \ee

Starting from $F_\lambda^{unabs}$ and applying the four prescriptions
for dust attenuation, we estimate the corresponding \lir values. We
compare them with \lirgs in fig.~\ref{fig:figL}: in each case
we see reasonable agreement between the predicted total amounts of
absorbed starlight and {\gs} results. This encouraging result is not
obvious, since our prescriptions assumed a fixed shape for the
extinction law (even if we consider an age-dependent component). This
implies a fixed ratio between the attenuation at different
wavelengths. On the other hand, we showed in Paper I (fig.~3) that the
scatter in {\gs} predicted attenuation laws is significant, and
although the Milky Way attenuation law is a good estimate for the mean
shape, single objects may deviate considerably. The agreement is
especially good for the disc-dominated objects, where we see a tight
relation between the stellar mass and \lirgs. The intrinsic
scatter increases in the bulge-dominated objects and the analytic
prescription are able to reproduce the general trend, even if the
residuals of the relation (right panels) show a $0.5$dex scatter.
This result suggests that the deviation of synthetic SEDs from a Milky
Way attenuation law is larger in bulge-dominated objects.  Under the
hypothesis that all the absorbed energy is re-emitted at infrared
wavelengths, and assuming a set of IR templates as described above, it
is then possible to estimate the corresponding infrared SEDs.

\subsection{Comparing IR SEDs}
\label{sec:comp_seds}

In order to compare the predictions of the full {\gs} RT calculation
to the IR template approach we split the model galaxies in the low-z
and high-z samples into IR luminosity classes, according to their
\lirgs.  For each IR template library, we define luminosity classes by
sorting each model galaxy into the nearest \lir bin.  We renormalise
each {\gs} SED in our sample to the ratio between \lirgs and the \lir
of the bin. We then compute the statistical properties of the
synthetic SEDs for each class: we create a mean SED, by considering
the mean flux at each wavelength, and we define a scatter in the flux
distribution as a function of wavelength. For the CE01 templates we
consider both the contribution of direct starlight and dust emission,
whereas for the other libraries we consider only the contribution of
the dust emission. We consider the low-z and high-z samples
separately.

We compare the mean {\gs} SEDs to the corresponding IR templates, for
four representative values of \lir and three libraries in
fig.~\ref{fig:sed_mean}. The agreement between the mean SEDs
constructed from our samples and the IR templates is satisfactory in
most cases, but there are some significant discrepancies. In order to
perform the same comparison over a wider range of \lir, we consider
the full dynamical range probed by the R09 library in
fig.~\ref{fig:r09_sed_mean}. The most prominent difference between the
templates and the predictions of the {\gs} RT-solver concerns the
position of the peak of the thermal dust emission: the peak is shifted
to longer wavelengths in the {\gs} predictions relative to the IR
templates.  It is particularly evident in fig.~\ref{fig:r09_sed_mean}
that the discrepancy depends also on \lir. In order to get a better
insight into this problem, we compare the mean {\gs} SEDs for
different values of the total amount of energy emitted in the IR in
fig.~\ref{fig:lir_evo} (left panel). \lir has a strong effect on the
overall normalisation of the mean SEDs, but they show a quite similar
shape. In particular, the position of the IR peak does not depend on
\lir\ over the whole range studied, while it is evident from
fig.~\ref{fig:r09_sed_mean}, that it is changing significantly in the
templates. Another notable discrepancy is present in the Mid-Infrared
($8 \mu$m$ < \lambda < 20 \mu$m), where the highly uncertain
contribution of PAH emission is dominant. Finally, it is also worth
noting the strong discrepancy seen in most cases at the shortest
wavelengths ($\lambda < 8 \mu$m).  This is the region where direct
starlight emission from old stars is still significant in the galactic
SEDs and comparable to the dust thermal emission: therefore we
interpret this result as due to the difficulties in disentangling the
relative contribution of these two components.

The discrepancy between the IR templates and the mean SEDs for the
low-z sample is particularly interesting, since the former have been
calibrated to the properties of local galaxies, and since {\gs} is
able to reproduce their SEDs, under a reasonable choice for dust
parameters \citep{Silva98}. In particular, the position depends mainly
on the temperature distribution of the emitting grains, and it is
likely connected to the assumed modeling of dust abundance and
distribution. This discrepancy is due to the combination of two
separate and degenerate effects. First of all, we assume the same dust
properties for all galaxies in our samples. This represents a common
choice in the semi-analytic framework, and it provides a reasonable
description for large model galaxy samples. Second, our synthetic SEDs
depend on the {\gal} predicted physical conditions of model galaxies;
we already know from previous work \citep{Fontanot07b, Kimm08} that
this particular {\gal} realization does not reproduce all properties
of local galaxies, i.e. their levels of star formation and their
relative gas and stellar content. Most notably, in the original
formulation of the model we do not attempt to calibrate our prediction
to reproduce the IR properties of local galaxies. Therefore it is
perfectly plausible to obtain mean SEDs which differ from the local
sample. Nonetheless, the proposed comparison is instructive regarding
the different results obtained by the two approaches for the generic
galaxy sample extracted from a cosmological galaxy formation model.

The most interesting insight from the analysis of
fig.~\ref{fig:sed_mean} and~\ref{fig:r09_sed_mean} is the different
shapes of the mean SEDs drawn from the high-z and low-z sample. In
most cases the mean SEDs corresponding to the high-z sample show on
average a better agreement with the corresponding template. This
result is somewhat surprising, given the fact that IR templates are
calibrated using low-z observations. In order to better understand
this result, we construct the mean SEDs over a finer binning in
redshift, while retaining the same \lir binning as for the R09
templates. We compare the evolution of the new mean SED sample with
redshift in fig.~\ref{fig:lir_evo} (right panel), for the
representative luminosity bin centred at $Log($\lir$)=10.5$.
Different luminosity classes show similar results. It is evident from
this plot that the mean SED is experiencing a strong redshift
evolution, in particular in the spectral regions corresponding to the
peak of thermal IR emission and PAH emission. This pattern can be
ascribed to the redshift evolution of the physical properties
(i.e. gas content, metal enrichment and size evolution) of the model
galaxies in the {\gal} realization and represents a warning against
using the same template library over a wide redshift range (as already
noticed by DN02 and R09). It is also worth mentioning that the same
dependence of the mean SED on \lir\ at fixed redshift
(fig.~\ref{fig:lir_evo}, left panel) seen at $z=0$ holds at higher
redshifts. In order to get better insight into the main driver of the
evolution of the SED shape, we perform a fitting procedure similar to
Paper I (see their sec.~4). For each synthetic SED in our library, we
modeled the wavelength corresponding to the peak of IR emission as a
power-law function of the physical properties of the corresponding
model galaxy. We consider \lir, the stellar and cold gas masses, the
star formation rate, metallicity and galactic radius and we define a
set of general relations involving independent quantities. For each
combination, we determine the best-fit parameters and evaluate the
goodness of the fit through a $\chi^2$ procedure. Our results show
that the strongest correlation is found for the total mass surface
density of the galaxy. This is not completely unexpected since the
relative spatial distribution of stars and gas is fundamental to
predicting the temperature of dust grains. The scatter in the
correlation is large, but we find that adding additional degrees of
freedom does not reduce it considerably. We conclude that other
physical quantities, like star formation rate and metallicity, still
play a non-negligible role in determining the resulting shape of the
SED.
\begin{figure*}
  \centerline{ 
    \includegraphics[width=9cm]{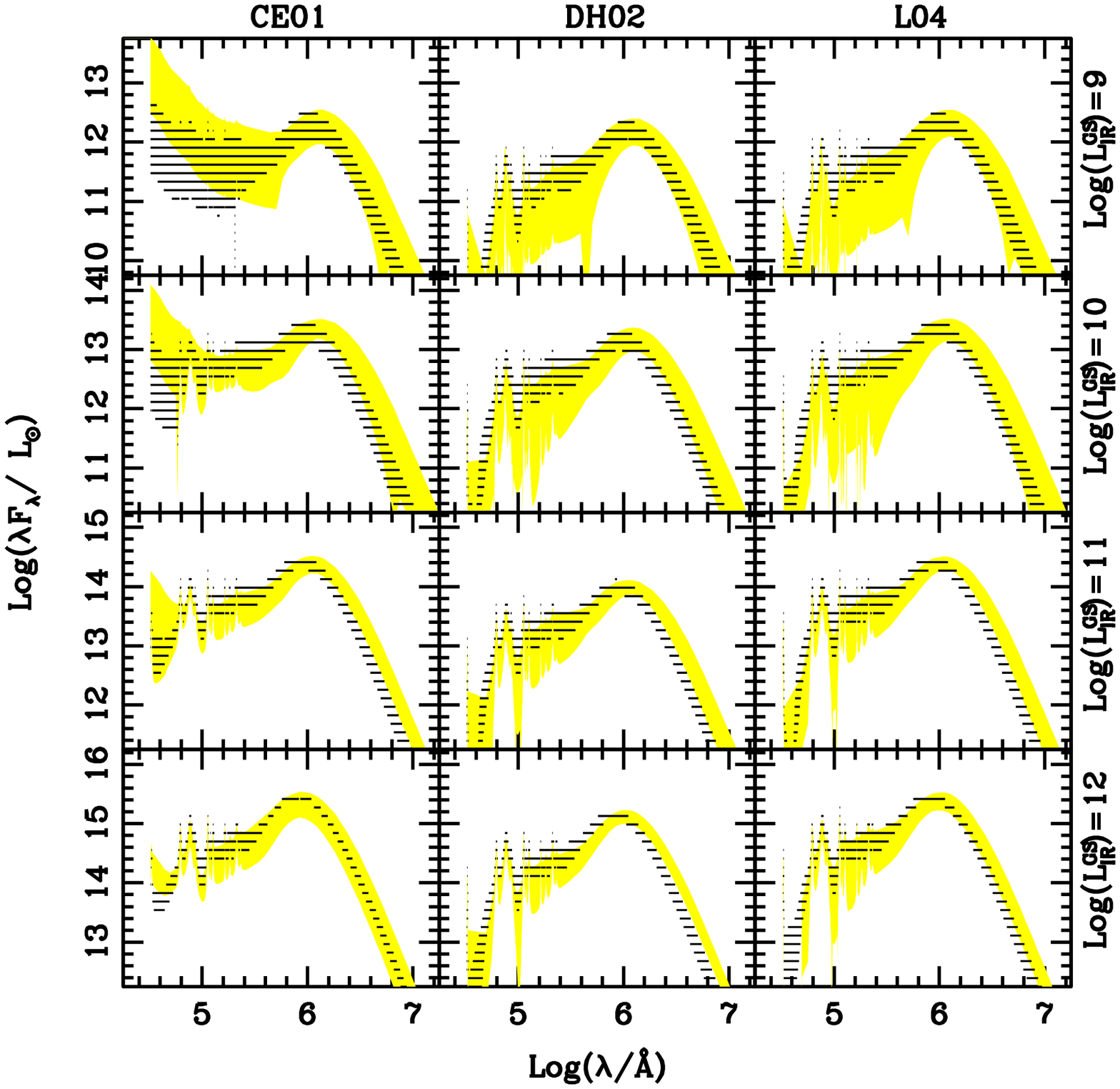} 
    \includegraphics[width=9cm]{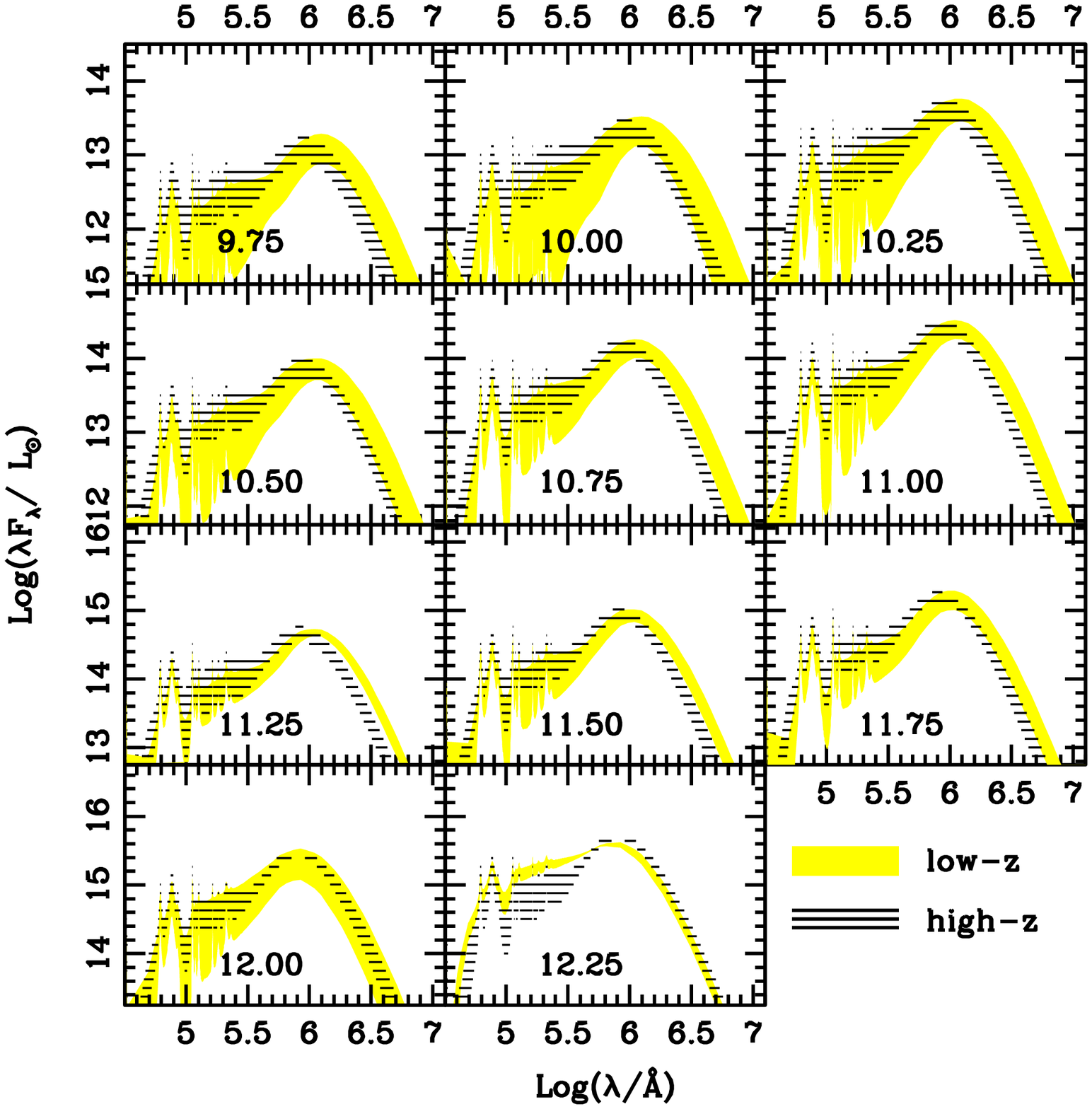} 
  }
  \caption{{\it Left Panels:} The 1-$\sigma$ distribution of {\gs}
    predicted SEDs in the same \lir reference intervals and for the
    same libraries as in fig~\ref{fig:sed_mean}. {\it Right Panels:}
    The 1-$\sigma$ distribution of {\gs} predicted SEDs for the same
    \lir reference intervals as defined in R09. The central \lir value
    for each interval is indicated in each panel. In all panels the
    yellow shaded area and horizontal texture refer to the low-z and
    high-z sample respectively.}
  \label{fig:sed_scatter}
\end{figure*}

We then consider the distribution of fluxes as a function of
wavelength in fig.~\ref{fig:sed_scatter}. It is evident from the plot
that the scatter within each \lir class may be important, especially
at the lowest values of total absorbed starlight. Both the scatter
variations between different libraries at a given \lir (left panels)
and along the same library (right panels) are mainly due to the
inhomogeneous distribution of galaxies in the \lir bins.  As expected,
larger binnings correspond to larger scatters around the mean SEDs. At
the same time, the variance of the flux as a function of wavelength
does not show any significant dependence on redshift, for a given \lir
bin. This result implies that using a single SED from a template
library to describe all galaxies belonging to a given \lir class, may
introduce a bias between the comparison between model predictions and
observations. Our analysis shows that it is possible to use mean SEDs
as representative for all SEDs belonging to a luminosity class only in
a statistical sense (and this proxy improves with finer binning in
\lir), but not on a object-by-object basis. It is also worth noting
that the different binning has a non-negligible effect on the shape
and normalisation of the corresponding mean SED, due to the different
sample definition. Although the IR templates share a similar trend,
this is another hint that a finer binning helps reduce the differences
between templates and {\gs} RT predictions.

\subsection{Effect on the IR Luminosity functions}
\label{sec:irlf}
\begin{figure}
  \centerline{
    \includegraphics[width=9cm]{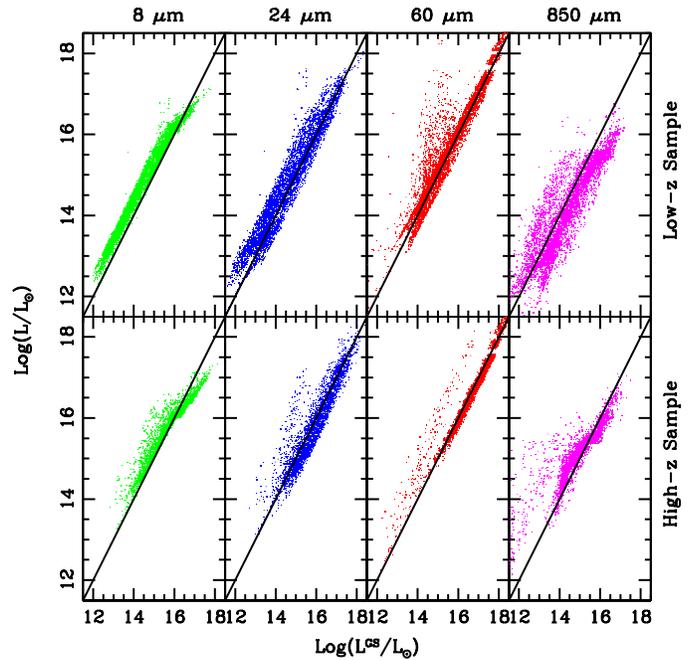}
  }
  \caption{Intrinsic IR luminosities at $8$, $24$, $60$, $850$ $\mu$m
    compared with the corresponding predictions for the R09 templates.}
  \label{fig:irflux}
\end{figure}

We then compare the individual monochromatic luminosities for each
model galaxy as predicted by the {\gs} computation and by the IR
templates. We select four wavelengths widely used in the literature to
track different galaxy populations and constrain the properties of
dust and PAHs at different temperatures: namely the $8 \mu m$, the $24
\mu m$, the $60 \mu m$ and the $850 \mu m$ restframe luminosities. For
each model galaxy we estimate the expected fluxes from the IR template
library as follows. We use the intrinsic {\gs}-predicted \lirgs
to select the closest template for each IR library. We then rescale
the template to the ratio between its \lir and \lirgs, and
we use it to estimate the monochromatic luminosities in the four
bands. For the CE01 templates we simply use the template, while for
the other libraries we sum up the chosen template with the
extinguished starlight spectrum (as predicted by {\gs}). We then
statistically compare the predictions derived using the template
libraries with the intrinsic results of {\gs} computations by
constructing the corresponding luminosity functions. In this way we
are able to quantify the impact of the different choice for IR
templates on the statistical properties of galaxy samples, as
predicted by SAMs. It is worth stressing here that we do not require
our SAM to reproduce the properties of the local Universe. Instead our
aim is to check the impact of different prescriptions for dust
emission on the predictions of SAMs.

\begin{figure*}
  \centerline{
    \includegraphics[width=15cm]{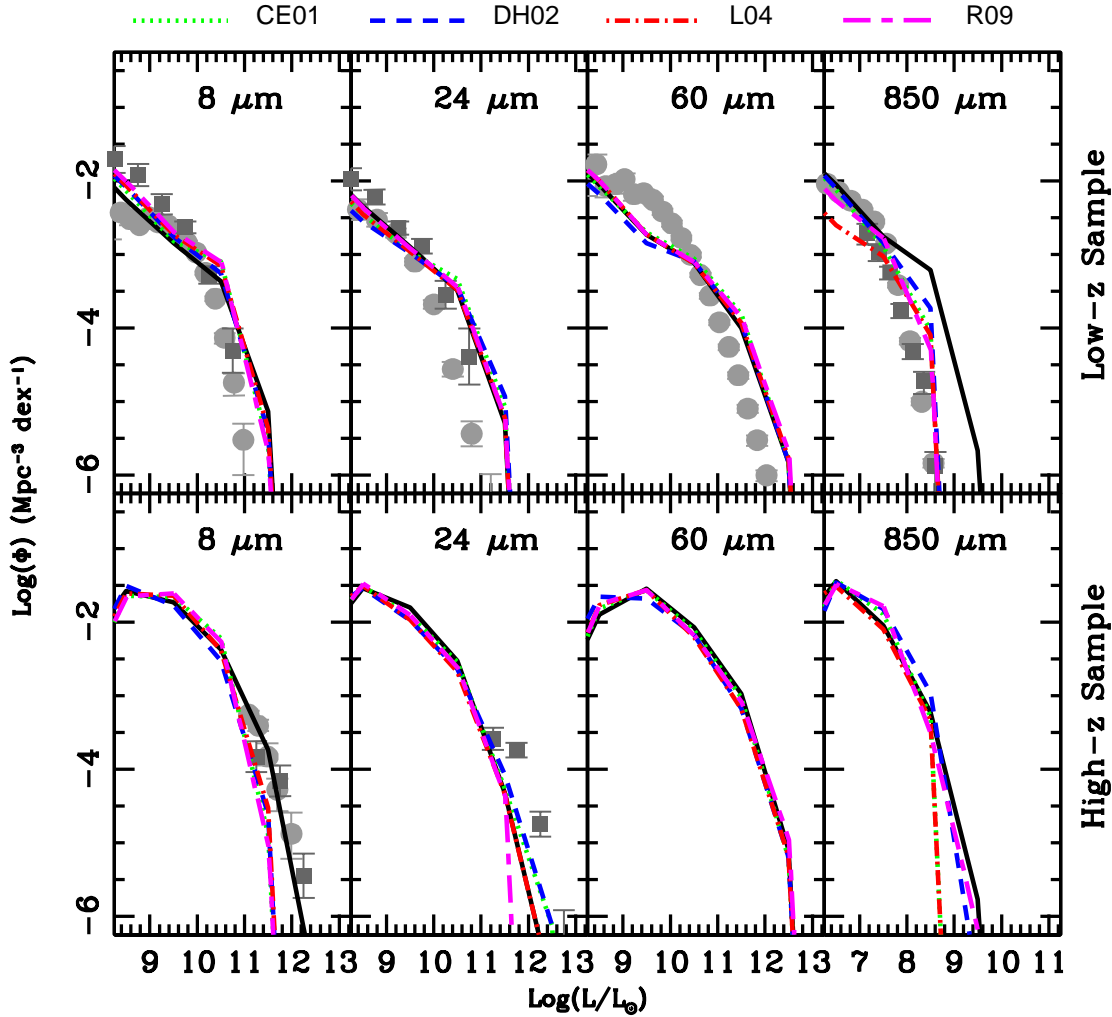}
  }
  \caption{{\gs} predicted IR luminosity functions in the low-z and
    high-z samples (thick solid lines) compared with the predictions
    using the IR template libraries. Solid, dotted, dashed, dot-dashed
    and long-short dashed lines refer to {\gs}, CE01, DH02, L04 and
    R09 templates respectively. Data points represent a compilation of
    local measurements from \citet[8 $\mu m$]{Huang07}, \citet[8 $\mu
      m$ and 24 $\mu m$]{Rodighiero10a} \citet[60 $\mu
      m$]{Saunders90}, \citet[850 $\mu m$]{Dunne00}, \citet[850 $\mu
      m$]{SerjeantHarrison05} and high redshift results from \citet[8
      $\mu m$]{Caputi07}, \citet[8 $\mu m$ and 24 $\mu
      m$]{Rodighiero10a} \citet[24 $\mu m$]{Marleau07}.}
  \label{fig:irlf}
\end{figure*}

In fig.~\ref{fig:irflux} we show a comparison between the
monochromatic luminosities obtained using the R09 templates with the
intrinsic {\gs} predictions (similar results hold for the other
templates), while in fig.~\ref{fig:irlf} the resulting luminosity
function for all considered templates (right panel). The intrinsic
luminosity functions for our low-z and high-z samples are represented
by the solid thick line, while other lines correspond to the
predictions obtained using the four IR template libraries. We find
similar results with respect to the analogous analysis at optical and
near-infrared wavelengths presented in Paper I (their fig.~12): both
the shape and normalisation of the luminosity functions are correctly
recovered in most cases. These results clearly show that the use of a
statistical estimator like the luminosity function, defined over a
large galaxy sample and including the contribution of various
luminosity classes, is less sensitive to uncertainties in the
prediction of individual galaxy properties. Nonetheless, some
important discrepancies are seen. In particular, in the sub-mm region,
{\gs} fluxes are systematically underpredicted by the templates: this
is mainly due to the discrepancy in the position of the $\sim 100
\mu$m peak in the SEDs (see fig.~\ref{fig:sed_mean}), which affects
the shape of the SEDs at longer wavelengths. In the low-z sample, the
flux at $850 \mu$m is underpredicted only for the brightest sources,
while at fainter fluxes the agreement is reasonable. However, in the
high-z sample the templates are not able to reproduce the brightest
sources at $8 \mu$m, while the agreement at $24 \mu$m and $60 \mu$m is
again satisfactory. It is possible to understand this peculiar
behaviour by looking at the shape of the brightest mean SEDs compared
with the corresponding templates in fig.~\ref{fig:sed_mean}. In order
to assess the impact of these effects in terms of the comparison
between model predictions and data, we show in fig.~\ref{fig:irlf} the
available observational constraints for the corresponding luminosity
functions. We warn the reader that we do not explicitly tune our model
to reproduce the corresponding samples; moreover, real data samples do
not always match the same redshift interval we adopt to define our SED
libraries. Nonetheless, we include them to guide the eye and to
provide an hint of the effect of different IR modeling in the
prediction of the luminosity function. As expected, the most
interesting results are seen for the 8 $\mu m$ and 850 $\mu m$
bands. In particular, the 8 $\mu m$ $z\sim2$ luminosity function is
better reproduced by the {\gs} RT predictions, while it is
underpredicted by IR template approach. On the other hand, the low-z
850 $\mu m$ luminosity function, which is clearly overpredicted in the
{\gal}+{\gs} framework, is recovered by the same galaxy formation
model when the IR template approach is adopted.

\section{Summary}
\label{sec:final}

This paper is the second in a series aimed at (i) understanding and
quantifying the results of detailed calculations of radiative transfer
in a dusty medium and (ii) comparing these results with simpler
analytic recipes coupled with semi-analytic models and
spectro-photometric codes. The final goal of this work is to assess
the impact of different modeling for dust attenuation and emission in
the framework of semi-analytic models of galaxy formation and
evolution. In the first paper of the series (Paper I) we studied the
effects of dust attenuation at optical and UV wavelengths, while in
this paper we focus on the dust emission in the IR region. We use the
same star formation history libraries based on the {\gal}
semi-analytic model and defined in Paper I: they include a low-z
($0<z<0.2$) and a high-z ($2<z<3$) sample, with the aim of assessing
the effect of the evolution of galaxy physical properties on the
corresponding SEDs. We couple the star formation histories and galaxy
properties to the RT-solver {\gs}, to obtain an estimate of the
synthetic SEDs from the UV to the Radio.

As a first check we consider the intrinsic absorbed starlight \lir
according to the {\gs} results and we compare it to the predictions of
the four prescriptions for dust attenuation at optical to
near-infrared wavelengths discussed in Paper I. As already shown in
Paper I, the scatter in the {\gs} predicted attenuation laws is not
negligible: therefore we expect discrepancies on an object-by-object
basis. However, our results show that all prescriptions provide a
reasonable description of the mean intrinsic \lir for the whole
population, within the errors shown in fig.~\ref{fig:figL}. We
conclude that such simplified analytic approaches can be used to
obtain statistically valid estimates for \lir.

We then consider four different IR template libraries, commonly
coupled to spectro-photometric models to provide predictions for dust
emission in the IR. They include both theoretical \citep{Dale02,
  Lagache04} and empirical templates \citep{Chary01, Rieke09}. In each
case, we define \lir classes similar to those used to generate the IR
templates and we sort the synthetic SEDs for the low-z and high-z
samples accordingly. For each class, we then construct mean SEDs by
computing the mean fluxes as a function of wavelength. Our results
show that, overall, the mean SEDs are in fairly good agreement with IR
templates over a wide spectral range. Significant discrepancies arise,
however, both at wavelengths longer than the $\sim 100 \mu$m peak, due
to the negligible evolution of the peak position with respect to the
IR templates, and for low IR luminosities. These discrepancies are
mainly due to the fact that the {\gal}+{\gs} code has not been
explicitly calibrated to reproduce the details of dust and physical
properties of IR bright galaxies in the local Universe, but it assumes
mean values for the relevant quantities.

Moreover, by comparing a low-z and a high-z sample we see significant
differences in the properties of mean SEDs, again growing larger for
low values of \lir. In order to gain better insight into the origin of
these discrepancies, we compute mean SEDs on a finer binning in the
\lir\ and redshift space.  We then consider the redshift evolution of
the mean SED at a fixed \lir\ (and vice versa). The mean SED shape
shows strong redshift evolution at fixed \lir, and very little
evolution as a function of \lir\ at a given cosmic epoch. We quantify
the dependence of the SED shape on the physical properties of the
underlying model galaxies by fitting the position of the peak of IR
emission. Our results show that the main contributor is the redshift
evolution of the spatial distribution of stars and gas (total mass
surface density) as predicted by {\gal}. Both star formation rate and
metallicity cause a secondary, but non-negligible contribution to the
broad scatter in the relation. This result reflects the fact that the
main factor responsible for the detailed shape of the {\gs} predicted
IR SED is the mean temperature of dust grains, which is a complex
function of the distribution of dust particles (linked to the cold gas
surface density and metallicity) and of the radiation field (linked to
the stellar mass surface density and star formation rate). Finally, we
also show that the scatter around the mean fluxes is significant and
it depends on the width of the \lir\ binning. For a sparse sampling of
\lir\ the mean SEDs give only a poor description of the SED variety.

These results suggest that it may not be valid to use the same IR
template library for modeling galaxies at different redshifts, where
their physical properties are likely different (as in the case of
{\gal} galaxies). Moreover, despite the fact that many physical
properties of galaxies (such as gas and metal content, present and
past star formation activity, disc and bulge sizes) are predicted by
the SAM, when solving the equation of RT we still have to make
assumptions about the physical state of the dust (composition, grain
size distribution) and about its distribution relative to stars and
the interstellar medium. Given the present uncertainties, we assume
that these properties do not change as a function of galactic
properties nor with cosmic time over the redshift range we
consider. Our choice reflects a parameter combination that has been
proven adequate for reproducing the properties of $z < 3$ galaxies
\citep{Silva98, Fontanot07b}.  However, a strong systematic variation
of these parameters (especially $t_{esc}$ and $f_{MC}$) at higher
redshifts is expected both observationally \citep{Maiolino04} and
theoretically \citep{LoFaro09}. We expect any variation in the dust
parameters and/or composition \citep{Schurer09} to lead to even larger
differences in the mean SED shape and normalisation.

We then consider the IR monochromatic luminosities at $8 \mu m$, $24
\mu m$, $60 \mu m$ and $850 \mu m$ as predicted by the {\gal}+{\gs}
model and by the four IR template libraries. We compare the fluxes on
an object-by-object basis and the resulting luminosity
functions. Although there are large discrepancies for some individual
objects (up to two orders of magnitude), the overall statistical
agreement is quite good in most regimes. There is a systematic
discrepancy at $850 \mu m$ for both the low-z and high-z samples, with
a systematic underprediction of the {\gs} RT predicted luminosity
functions. This is mainly related to the discrepant position of the
$\sim 100 \mu$m peak in the SEDs with respect to the templates. The
brightest sources in the high-z sample show disagreement also at $8
\mu$m, while the $24 \mu$m and $60 \mu$m predictions are similar. This
is related to the peculiar shape of the high-z sample mean SED at this
IR luminosity, and, again, implies that caution should be used when
comparing theoretical IR predictions from different methods, even on a
statistical basis.

Our results extend those presented in Paper I into the IR regime, and
our conclusions are quite similar. The level of agreement between the
predictions of a RT solver approach (superior in order to understand
the details of dust absorption and re-emission in galaxies on a
object-by-object basis) and more computationally efficient
semi-analytic methods depends on the spectral regions under
analysis. We have identified wavebands where the agreement is good (in
a statistical sense) and others where the discrepancies are
significant. In particular, for the {\gal}+{\gs} model the agreement
between the two approaches is good in the Mid-IR, while we found
significant discrepancies at 8$\mu$m and 850$\mu$m.

As a wealth of multi-wavelength observations are becoming available in
the IR, we expect to use them to constrain with greater accuracy the
redshift evolution of dust properties and their relation to the
physical properties of galaxies, as well as the observed shape of
their SEDs over a wide range of redshift. It has been recently shown
that the computational time required for a full RT solver coupled to a
SAM can be considerably reduced, with no loss of accuracy, with the
use of neural networks \citep{Silva10}, thus helping to overcome one
of the biggest drawbacks connected to the use of this tool in the SAM
framework. At the same time, improvements in the SAMs are needed,
i.e. the modeling of dust generation, evolution and dispersion, in
order to reduce the number of externally fixed parameters and make use
of the full predictive power of RT solvers.

\section*{Acknowledgments}
The authors would like to thank Pierluigi Monaco and Gabriella de
Lucia for stimulating discussions. Some of the calculations were
carried out on the PIA cluster of the Max-Planck-Institut f\"ur
Astronomie at the Rechenzentrum Garching. FF acknowledges the support
of an INAF-OATs fellowship granted on 'Basic Research' funds and
hospitality at the Kavli Institute for Theoretical Physics in Santa
Barbara. This research was supported in part by the National Science
Foundation under Grant No. NSF PHY05-51164.
\bibliographystyle{mn2e} 
\bibliography{fontanot}

\end{document}


%% file: ir_model.bbl
\begin{thebibliography}{}

\bibitem[\protect\citeauthoryear{{Babbedge}, {Rowan-Robinson}, {Vaccari},
  {Surace}, {Lonsdale}, {Clements}, {Fang} \& {Farrah}}{{Babbedge}
  et~al.}{2006}]{Babbedge06}
{Babbedge} T.~S.~R.,  {Rowan-Robinson} M.,  {Vaccari} M.,  {Surace} J.~A.,
  {Lonsdale} C.~J.,  {Clements} D.~L.,  {Fang} F.,    {Farrah} D. e.~a.,  2006,
  \mnras, 370, 1159

\bibitem[\protect\citeauthoryear{{Baugh}, {Lacey}, {Frenk}, {Granato}, {Silva},
  {Bressan}, {Benson} \& {Cole}}{{Baugh} et~al.}{2005}]{Baugh05}
{Baugh} C.~M.,  {Lacey} C.~G.,  {Frenk} C.~S.,  {Granato} G.~L.,  {Silva} L.,
  {Bressan} A.,  {Benson} A.~J.,    {Cole} S.,  2005, \mnras, 356, 1191

\bibitem[\protect\citeauthoryear{{Bouwens}, {Illingworth}, {Franx} \&
  {Ford}}{{Bouwens} et~al.}{2007}]{Bouwens07}
{Bouwens} R.~J.,  {Illingworth} G.~D.,  {Franx} M.,    {Ford} H.,  2007, \apj,
  670, 928

\bibitem[\protect\citeauthoryear{{Bower}, {Benson}, {Malbon}, {Helly}, {Frenk},
  {Baugh}, {Cole} \& {Lacey}}{{Bower} et~al.}{2006}]{Bower06}
{Bower} R.~G.,  {Benson} A.~J.,  {Malbon} R.,  {Helly} J.~C.,  {Frenk} C.~S.,
  {Baugh} C.~M.,  {Cole} S.,    {Lacey} C.~G.,  2006, \mnras, 370, 645

\bibitem[\protect\citeauthoryear{{Bressan}, {Granato} \& {Silva}}{{Bressan}
  et~al.}{1998}]{Bressan98}
{Bressan} A.,  {Granato} G.~L.,    {Silva} L.,  1998, \aap, 332, 135

\bibitem[\protect\citeauthoryear{{Bressan}, {Silva} \& {Granato}}{{Bressan}
  et~al.}{2002}]{Bressan02}
{Bressan} A.,  {Silva} L.,    {Granato} G.~L.,  2002, \aap, 392, 377

\bibitem[\protect\citeauthoryear{{Bruzual} \& {Charlot}}{{Bruzual} \&
  {Charlot}}{2003}]{Bruzual03}
{Bruzual} G.,  {Charlot} S.,  2003, \mnras, 344, 1000

\bibitem[\protect\citeauthoryear{{Calzetti}, {Armus}, {Bohlin}, {Kinney},
  {Koornneef} \& {Storchi-Bergmann}}{{Calzetti} et~al.}{2000}]{Calzetti00}
{Calzetti} D.,  {Armus} L.,  {Bohlin} R.~C.,  {Kinney} A.~L.,  {Koornneef} J.,
    {Storchi-Bergmann} T.,  2000, \apj, 533, 682

\bibitem[\protect\citeauthoryear{{Caputi}, {Lagache}, {Yan}, {Dole},
  {Bavouzet}, {Le Floc'h}, {Choi}, {Helou} \& {Reddy}}{{Caputi}
  et~al.}{2007}]{Caputi07}
{Caputi} K.~I.,  {Lagache} G.,  {Yan} L.,  {Dole} H.,  {Bavouzet} N.,  {Le
  Floc'h} E.,  {Choi} P.~I.,  {Helou} G.,    {Reddy} N.,  2007, \apj, 660, 97

\bibitem[\protect\citeauthoryear{{Chapman}, {Blain}, {Ivison} \&
  {Smail}}{{Chapman} et~al.}{2003}]{Chapman03}
{Chapman} S.~C.,  {Blain} A.~W.,  {Ivison} R.~J.,    {Smail} I.~R.,  2003,
  \nat, 422, 695

\bibitem[\protect\citeauthoryear{{Chapman}, {Blain}, {Smail} \&
  {Ivison}}{{Chapman} et~al.}{2005}]{Chapman05}
{Chapman} S.~C.,  {Blain} A.~W.,  {Smail} I.,    {Ivison} R.~J.,  2005, \apj,
  622, 772

\bibitem[\protect\citeauthoryear{{Charlot} \& {Fall}}{{Charlot} \&
  {Fall}}{2000}]{CharlotFall00}
{Charlot} S.,  {Fall} S.~M.,  2000, \apj, 539, 718

\bibitem[\protect\citeauthoryear{{Chary} \& {Elbaz}}{{Chary} \&
  {Elbaz}}{2001}]{Chary01}
{Chary} R.,  {Elbaz} D.,  2001, \apj, 556, 562

\bibitem[\protect\citeauthoryear{{Cole}, {Lacey}, {Baugh} \& {Frenk}}{{Cole}
  et~al.}{2000}]{Cole00}
{Cole} S.,  {Lacey} C.~G.,  {Baugh} C.~M.,    {Frenk} C.~S.,  2000, \mnras,
  319, 168

\bibitem[\protect\citeauthoryear{{Croton}, {Springel}, {White}, {De Lucia},
  {Frenk}, {Gao}, {Jenkins}, {Kauffmann}, {Navarro} \& {Yoshida}}{{Croton}
  et~al.}{2006}]{Croton06}
{Croton} D.~J.,  {Springel} V.,  {White} S.~D.~M.,  {De Lucia} G.,  {Frenk}
  C.~S.,  {Gao} L.,  {Jenkins} A.,  {Kauffmann} G.,  {Navarro} J.~F.,
  {Yoshida} N.,  2006, \mnras, 365, 11

\bibitem[\protect\citeauthoryear{{Dale} \& {Helou}}{{Dale} \&
  {Helou}}{2002}]{Dale02}
{Dale} D.~A.,  {Helou} G.,  2002, \apj, 576, 159

\bibitem[\protect\citeauthoryear{{Dale}, {Helou}, {Contursi}, {Silbermann} \&
  {Kolhatkar}}{{Dale} et~al.}{2001}]{Dale01}
{Dale} D.~A.,  {Helou} G.,  {Contursi} A.,  {Silbermann} N.~A.,    {Kolhatkar}
  S.,  2001, \apj, 549, 215

\bibitem[\protect\citeauthoryear{{De Lucia} \& {Blaizot}}{{De Lucia} \&
  {Blaizot}}{2007}]{DeLucia07b}
{De Lucia} G.,  {Blaizot} J.,  2007, \mnras, 375, 2

\bibitem[\protect\citeauthoryear{{De Lucia}, {Springel}, {White}, {Croton} \&
  {Kauffmann}}{{De Lucia} et~al.}{2006}]{DeLucia06}
{De Lucia} G.,  {Springel} V.,  {White} S.~D.~M.,  {Croton} D.,    {Kauffmann}
  G.,  2006, \mnras, 366, 499

\bibitem[\protect\citeauthoryear{{Desert}, {Boulanger} \& {Puget}}{{Desert}
  et~al.}{1990}]{Desert90}
{Desert} F.-X.,  {Boulanger} F.,    {Puget} J.~L.,  1990, \aap, 237, 215

\bibitem[\protect\citeauthoryear{{Devriendt} \& {Guiderdoni}}{{Devriendt} \&
  {Guiderdoni}}{2000}]{Devriendt00}
{Devriendt} J.~E.~G.,  {Guiderdoni} B.,  2000, \aap, 363, 851

\bibitem[\protect\citeauthoryear{{Devriendt}, {Guiderdoni} \&
  {Sadat}}{{Devriendt} et~al.}{1999}]{Devriendt99}
{Devriendt} J.~E.~G.,  {Guiderdoni} B.,    {Sadat} R.,  1999, \aap, 350, 381

\bibitem[\protect\citeauthoryear{{Dole}, {Gispert}, {Lagache}, {Puget},
  {Bouchet}, {Cesarsky}, {Ciliegi} \& {Clements}}{{Dole} et~al.}{2001}]{Dole01}
{Dole} H.,  {Gispert} R.,  {Lagache} G.,  {Puget} J.-L.,  {Bouchet} F.~R.,
  {Cesarsky} C.,  {Ciliegi} P.,    {Clements} D.~L. e.~a.,  2001, \aap, 372,
  364

\bibitem[\protect\citeauthoryear{{Dopita}, {Groves}, {Fischera}, {Sutherland},
  {Tuffs}, {Popescu}, {Kewley}, {Reuland} \& {Leitherer}}{{Dopita}
  et~al.}{2005}]{Dopita05}
{Dopita} M.~A.,  {Groves} B.~A.,  {Fischera} J.,  {Sutherland} R.~S.,  {Tuffs}
  R.~J.,  {Popescu} C.~C.,  {Kewley} L.~J.,  {Reuland} M.,    {Leitherer} C.,
  2005, \apj, 619, 755

\bibitem[\protect\citeauthoryear{{Dorschner} \& {Henning}}{{Dorschner} \&
  {Henning}}{1995}]{DorschnerHenning95}
{Dorschner} J.,  {Henning} T.,  1995, \aapr, 6, 271

\bibitem[\protect\citeauthoryear{{Dunne}, {Eales}, {Edmunds}, {Ivison},
  {Alexander} \& {Clements}}{{Dunne} et~al.}{2000}]{Dunne00}
{Dunne} L.,  {Eales} S.,  {Edmunds} M.,  {Ivison} R.,  {Alexander} P.,
  {Clements} D.~L.,  2000, \mnras, 315, 115

\bibitem[\protect\citeauthoryear{{Elbaz}, {Cesarsky}, {Chanial}, {Aussel},
  {Franceschini}, {Fadda} \& {Chary}}{{Elbaz} et~al.}{2002}]{Elbaz02}
{Elbaz} D.,  {Cesarsky} C.~J.,  {Chanial} P.,  {Aussel} H.,  {Franceschini} A.,
   {Fadda} D.,    {Chary} R.~R.,  2002, \aap, 384, 848

\bibitem[\protect\citeauthoryear{{Elbaz}, {Cesarsky}, {Fadda}, {Aussel},
  {D{\'e}sert}, {Franceschini}, {Flores} \& {Harwit}}{{Elbaz}
  et~al.}{1999}]{Elbaz99}
{Elbaz} D.,  {Cesarsky} C.~J.,  {Fadda} D.,  {Aussel} H.,  {D{\'e}sert} F.~X.,
  {Franceschini} A.,  {Flores} H.,    {Harwit} M. e.~a.,  1999, \aap, 351, L37

\bibitem[\protect\citeauthoryear{{Fioc} \& {Rocca-Volmerange}}{{Fioc} \&
  {Rocca-Volmerange}}{1997}]{FiocRV97}
{Fioc} M.,  {Rocca-Volmerange} B.,  1997, \aap, 326, 950

\bibitem[\protect\citeauthoryear{{Fontana}, {Salimbeni}, {Grazian},
  {Giallongo}, {Pentericci}, {Nonino}, {Fontanot}, {Menci}, {Monaco},
  {Cristiani}, {Vanzella}, {de Santis} \& {Gallozzi}}{{Fontana}
  et~al.}{2006}]{Fontana06}
{Fontana} A.,  {Salimbeni} S.,  {Grazian} A.,  {Giallongo} E.,  {Pentericci}
  L.,  {Nonino} M.,  {Fontanot} F.,  {Menci} N.,  {Monaco} P.,  {Cristiani} S.,
   {Vanzella} E.,  {de Santis} C.,    {Gallozzi} S.,  2006, \aap, 459, 745

\bibitem[\protect\citeauthoryear{{Fontanot}, {De Lucia}, {Monaco}, {Somerville}
  \& {Santini}}{{Fontanot} et~al.}{2009}]{Fontanot09b}
{Fontanot} F.,  {De Lucia} G.,  {Monaco} P.,  {Somerville} R.~S.,    {Santini}
  P.,  2009, \mnras, 397, 1776

\bibitem[\protect\citeauthoryear{{Fontanot}, {Monaco}, {Cristiani} \&
  {Tozzi}}{{Fontanot} et~al.}{2006}]{Fontanot06}
{Fontanot} F.,  {Monaco} P.,  {Cristiani} S.,    {Tozzi} P.,  2006, \mnras,
  373, 1173

\bibitem[\protect\citeauthoryear{{Fontanot}, {Monaco}, {Silva} \&
  {Grazian}}{{Fontanot} et~al.}{2007}]{Fontanot07b}
{Fontanot} F.,  {Monaco} P.,  {Silva} L.,    {Grazian} A.,  2007, \mnras, 382,
  903

\bibitem[\protect\citeauthoryear{{Fontanot}, {Somerville}, {Silva}, {Monaco} \&
  {Skibba}}{{Fontanot} et~al.}{2009}]{Fontanot09a}
{Fontanot} F.,  {Somerville} R.~S.,  {Silva} L.,  {Monaco} P.,    {Skibba} R.,
  2009, \mnras, 392, 553

\bibitem[\protect\citeauthoryear{{Granato}, {Lacey}, {Silva}, {Bressan},
  {Baugh}, {Cole} \& {Frenk}}{{Granato} et~al.}{2000}]{Granato00}
{Granato} G.~L.,  {Lacey} C.~G.,  {Silva} L.,  {Bressan} A.,  {Baugh} C.~M.,
  {Cole} S.,    {Frenk} C.~S.,  2000, \apj, 542, 710

\bibitem[\protect\citeauthoryear{{Gruppioni}, {Lari}, {Pozzi}, {Zamorani},
  {Franceschini}, {Oliver}, {Rowan-Robinson} \& {Serjeant}}{{Gruppioni}
  et~al.}{2002}]{Gruppioni02}
{Gruppioni} C.,  {Lari} C.,  {Pozzi} F.,  {Zamorani} G.,  {Franceschini} A.,
  {Oliver} S.,  {Rowan-Robinson} M.,    {Serjeant} S.,  2002, \mnras, 335, 831

\bibitem[\protect\citeauthoryear{{Guiderdoni} \&
  {Rocca-Volmerange}}{{Guiderdoni} \& {Rocca-Volmerange}}{1987}]{GRV87}
{Guiderdoni} B.,  {Rocca-Volmerange} B.,  1987, \aap, 186, 1

\bibitem[\protect\citeauthoryear{{Hatton}, {Devriendt}, {Ninin}, {Bouchet},
  {Guiderdoni} \& {Vibert}}{{Hatton} et~al.}{2003}]{Hatton03}
{Hatton} S.,  {Devriendt} J.~E.~G.,  {Ninin} S.,  {Bouchet} F.~R.,
  {Guiderdoni} B.,    {Vibert} D.,  2003, \mnras, 343, 75

\bibitem[\protect\citeauthoryear{{Hauser} \& {Dwek}}{{Hauser} \&
  {Dwek}}{2001}]{HauserDwek01}
{Hauser} M.~G.,  {Dwek} E.,  2001, \araa, 39, 249

\bibitem[\protect\citeauthoryear{{Huang}, {Ashby}, {Barmby}, {Brodwin},
  {Brown}, {Caldwell}, {Cool} \& {Eisenhardt}}{{Huang} et~al.}{2007}]{Huang07}
{Huang} J.,  {Ashby} M.~L.~N.,  {Barmby} P.,  {Brodwin} M.,  {Brown} M.~J.~I.,
  {Caldwell} N.,  {Cool} R.~J.,    {Eisenhardt} e.~a.,  2007, \apj, 664, 840

\bibitem[\protect\citeauthoryear{{Jonsson}}{{Jonsson}}{2006}]{Jonsson06}
{Jonsson} P.,  2006, \mnras, 372, 2

\bibitem[\protect\citeauthoryear{{Kimm}, {Somerville}, {Yi}, {van den Bosch},
  {Salim}, {Fontanot}, {Monaco}, {Mo}, {Pasquali}, {Rich} \& {Yang}}{{Kimm}
  et~al.}{2009}]{Kimm08}
{Kimm} T.,  {Somerville} R.~S.,  {Yi} S.~K.,  {van den Bosch} F.~C.,  {Salim}
  S.,  {Fontanot} F.,  {Monaco} P.,  {Mo} H.,  {Pasquali} A.,  {Rich} R.~M.,
  {Yang} X.,  2009, \mnras, 394, 1131

\bibitem[\protect\citeauthoryear{{Lacey}, {Baugh}, {Frenk}, {Silva}, {Granato}
  \& {Bressan}}{{Lacey} et~al.}{2008}]{Lacey08}
{Lacey} C.~G.,  {Baugh} C.~M.,  {Frenk} C.~S.,  {Silva} L.,  {Granato} G.~L.,
   {Bressan} A.,  2008, \mnras, 385, 1155

\bibitem[\protect\citeauthoryear{{Lagache}, {Abergel}, {Boulanger},
  {D{\'e}sert} \& {Puget}}{{Lagache} et~al.}{1999}]{Lagache99}
{Lagache} G.,  {Abergel} A.,  {Boulanger} F.,  {D{\'e}sert} F.~X.,    {Puget}
  J.-L.,  1999, \aap, 344, 322

\bibitem[\protect\citeauthoryear{{Lagache}, {Dole} \& {Puget}}{{Lagache}
  et~al.}{2003}]{Lagache03}
{Lagache} G.,  {Dole} H.,    {Puget} J.-L.,  2003, \mnras, 338, 555

\bibitem[\protect\citeauthoryear{{Lagache}, {Dole}, {Puget},
  {P{\'e}rez-Gonz{\'a}lez}, {Le Floc'h}, {Rieke}, {Papovich}, {Egami},
  {Alonso-Herrero}, {Engelbracht}, {Gordon}, {Misselt} \& {Morrison}}{{Lagache}
  et~al.}{2004}]{Lagache04}
{Lagache} G.,  {Dole} H.,  {Puget} J.-L.,  {P{\'e}rez-Gonz{\'a}lez} P.~G.,  {Le
  Floc'h} E.,  {Rieke} G.~H.,  {Papovich} C.,  {Egami} E.,  {Alonso-Herrero}
  A.,  {Engelbracht} C.~W.,  {Gordon} K.~D.,  {Misselt} K.~A.,    {Morrison}
  J.~E.,  2004, \apjs, 154, 112

\bibitem[\protect\citeauthoryear{{Le Floc'h}, {Papovich}, {Dole}, {Bell},
  {Lagache}, {Rieke}, {Egami} \& {P{\'e}rez-Gonz{\'a}lez}}{{Le Floc'h}
  et~al.}{2005}]{LeFloch05}
{Le Floc'h} E.,  {Papovich} C.,  {Dole} H.,  {Bell} E.~F.,  {Lagache} G.,
  {Rieke} G.~H.,  {Egami} E.,    {P{\'e}rez-Gonz{\'a}lez} P.~G. e.~a.,  2005,
  \apj, 632, 169

\bibitem[\protect\citeauthoryear{{Leitherer}, {Schaerer}, {Goldader},
  {Delgado}, {Robert}, {Kune}, {de Mello}, {Devost} \& {Heckman}}{{Leitherer}
  et~al.}{1999}]{Leitherer99}
{Leitherer} C.,  {Schaerer} D.,  {Goldader} J.~D.,  {Delgado} R.~M.~G.,
  {Robert} C.,  {Kune} D.~F.,  {de Mello} D.~F.,  {Devost} D.,    {Heckman}
  T.~M.,  1999, \apjs, 123, 3

\bibitem[\protect\citeauthoryear{{Li} \& {Draine}}{{Li} \&
  {Draine}}{2001}]{LiDraine01}
{Li} A.,  {Draine} B.~T.,  2001, \apj, 554, 778

\bibitem[\protect\citeauthoryear{{Lo Faro}, {Monaco}, {Vanzella}, {Fontanot},
  {Silva} \& {Cristiani}}{{Lo Faro} et~al.}{2009}]{LoFaro09}
{Lo Faro} B.,  {Monaco} P.,  {Vanzella} E.,  {Fontanot} F.,  {Silva} L.,
  {Cristiani} S.,  2009, \mnras, 399, 827

\bibitem[\protect\citeauthoryear{{Maiolino}, {Schneider}, {Oliva}, {Bianchi},
  {Ferrara}, {Mannucci}, {Pedani} \& {Roca Sogorb}}{{Maiolino}
  et~al.}{2004}]{Maiolino04}
{Maiolino} R.,  {Schneider} R.,  {Oliva} E.,  {Bianchi} S.,  {Ferrara} A.,
  {Mannucci} F.,  {Pedani} M.,    {Roca Sogorb} M.,  2004, \nat, 431, 533

\bibitem[\protect\citeauthoryear{{Marleau}, {Fadda}, {Appleton},
  {Noriega-Crespo}, {Im} \& {Clancy}}{{Marleau} et~al.}{2007}]{Marleau07}
{Marleau} F.~R.,  {Fadda} D.,  {Appleton} P.~N.,  {Noriega-Crespo} A.,  {Im}
  M.,    {Clancy} D.,  2007, \apj, 663, 218

\bibitem[\protect\citeauthoryear{{Mathis}, {Mezger} \& {Panagia}}{{Mathis}
  et~al.}{1983}]{Mathis83}
{Mathis} J.~S.,  {Mezger} P.~G.,    {Panagia} N.,  1983, \aap, 128, 212

\bibitem[\protect\citeauthoryear{{Mo}, {Mao} \& {White}}{{Mo}
  et~al.}{1998}]{MoMaoWhite98}
{Mo} H.~J.,  {Mao} S.,    {White} S.~D.~M.,  1998, \mnras, 295, 319

\bibitem[\protect\citeauthoryear{{Monaco}}{{Monaco}}{2004}]{Monaco04}
{Monaco} P.,  2004, \mnras, 352, 181

\bibitem[\protect\citeauthoryear{{Monaco}, {Fontanot} \& {Taffoni}}{{Monaco}
  et~al.}{2007}]{Monaco07}
{Monaco} P.,  {Fontanot} F.,    {Taffoni} G.,  2007, \mnras, 375, 1189

\bibitem[\protect\citeauthoryear{{Monaco}, {Murante}, {Borgani} \&
  {Fontanot}}{{Monaco} et~al.}{2006}]{Monaco06}
{Monaco} P.,  {Murante} G.,  {Borgani} S.,    {Fontanot} F.,  2006, \apjl, 652,
  L89

\bibitem[\protect\citeauthoryear{{Panuzzo}, {Bressan}, {Granato}, {Silva} \&
  {Danese}}{{Panuzzo} et~al.}{2003}]{Panuzzo03}
{Panuzzo} P.,  {Bressan} A.,  {Granato} G.~L.,  {Silva} L.,    {Danese} L.,
  2003, \aap, 409, 99

\bibitem[\protect\citeauthoryear{{Panuzzo}, {Granato}, {Buat}, {Inoue},
  {Silva}, {Iglesias-P{\'a}ramo} \& {Bressan}}{{Panuzzo}
  et~al.}{2007}]{Panuzzo07}
{Panuzzo} P.,  {Granato} G.~L.,  {Buat} V.,  {Inoue} A.~K.,  {Silva} L.,
  {Iglesias-P{\'a}ramo} J.,    {Bressan} A.,  2007, \mnras, 375, 640

\bibitem[\protect\citeauthoryear{{Popescu}, {Misiriotis}, {Kylafis}, {Tuffs} \&
  {Fischera}}{{Popescu} et~al.}{2000}]{Popescu00}
{Popescu} C.~C.,  {Misiriotis} A.,  {Kylafis} N.~D.,  {Tuffs} R.~J.,
  {Fischera} J.,  2000, \aap, 362, 138

\bibitem[\protect\citeauthoryear{{Popescu} \& {Tuffs}}{{Popescu} \&
  {Tuffs}}{2002}]{Popescu02}
{Popescu} C.~C.,  {Tuffs} R.~J.,  2002, \mnras, 335, L41

\bibitem[\protect\citeauthoryear{{Pozzetti}, {Madau}, {Zamorani}, {Ferguson} \&
  {Bruzual A.}}{{Pozzetti} et~al.}{1998}]{Pozzetti98}
{Pozzetti} L.,  {Madau} P.,  {Zamorani} G.,  {Ferguson} H.~C.,    {Bruzual A.}
  G.,  1998, \mnras, 298, 1133

\bibitem[\protect\citeauthoryear{{Rieke}, {Alonso-Herrero}, {Weiner},
  {P{\'e}rez-Gonz{\'a}lez}, {Blaylock}, {Donley} \& {Marcillac}}{{Rieke}
  et~al.}{2009}]{Rieke09}
{Rieke} G.~H.,  {Alonso-Herrero} A.,  {Weiner} B.~J.,  {P{\'e}rez-Gonz{\'a}lez}
  P.~G.,  {Blaylock} M.,  {Donley} J.~L.,    {Marcillac} D.,  2009, \apj, 692,
  556

\bibitem[\protect\citeauthoryear{{Rodighiero}, {Vaccari}, {Franceschini},
  {Tresse}, {Le Fevre}, {Le Brun}, {Mancini} \& {Matute}}{{Rodighiero}
  et~al.}{2010}]{Rodighiero10a}
{Rodighiero} G.,  {Vaccari} M.,  {Franceschini} A.,  {Tresse} L.,  {Le Fevre}
  O.,  {Le Brun} V.,  {Mancini} C.,    {Matute} e.~a.,  2010, \aap, 515, A8+

\bibitem[\protect\citeauthoryear{{Salpeter}}{{Salpeter}}{1955}]{Salpeter55}
{Salpeter} E.~E.,  1955, \apj, 121, 161

\bibitem[\protect\citeauthoryear{{Sanders} \& {Mirabel}}{{Sanders} \&
  {Mirabel}}{1996}]{Sanders96}
{Sanders} D.~B.,  {Mirabel} I.~F.,  1996, \araa, 34, 749

\bibitem[\protect\citeauthoryear{{Saunders}, {Rowan-Robinson}, {Lawrence},
  {Efstathiou}, {Kaiser}, {Ellis} \& {Frenk}}{{Saunders}
  et~al.}{1990}]{Saunders90}
{Saunders} W.,  {Rowan-Robinson} M.,  {Lawrence} A.,  {Efstathiou} G.,
  {Kaiser} N.,  {Ellis} R.~S.,    {Frenk} C.~S.,  1990, \mnras, 242, 318

\bibitem[\protect\citeauthoryear{{Schurer}, {Calura}, {Silva}, {Pipino},
  {Granato}, {Matteucci} \& {Maiolino}}{{Schurer} et~al.}{2009}]{Schurer09}
{Schurer} A.,  {Calura} F.,  {Silva} L.,  {Pipino} A.,  {Granato} G.~L.,
  {Matteucci} F.,    {Maiolino} R.,  2009, \mnras, 394, 2001

\bibitem[\protect\citeauthoryear{{Serjeant} \& {Harrison}}{{Serjeant} \&
  {Harrison}}{2005}]{SerjeantHarrison05}
{Serjeant} S.,  {Harrison} D.,  2005, \mnras, 356, 192

\bibitem[\protect\citeauthoryear{{Silva}}{{Silva}}{1999}]{Silva99}
{Silva} L.,  1999, PhD thesis, AA(SISSA - Trieste (Italy))

\bibitem[\protect\citeauthoryear{{Silva}, {De Zotti}, {Granato}, {Maiolino} \&
  {Danese}}{{Silva} et~al.}{2005}]{Silva05}
{Silva} L.,  {De Zotti} G.,  {Granato} G.~L.,  {Maiolino} R.,    {Danese} L.,
  2005, \mnras, 357, 1295

\bibitem[\protect\citeauthoryear{{Silva}, {Granato}, {Bressan} \&
  {Danese}}{{Silva} et~al.}{1998}]{Silva98}
{Silva} L.,  {Granato} G.~L.,  {Bressan} A.,    {Danese} L.,  1998, \apj, 509,
  103

\bibitem[\protect\citeauthoryear{{Silva}, {Schurer}, {Granato}, {Almeida},
  {Baugh}, {Frenk}, {Lacey}, {Paoletti}, {Petrella} \& {Selvestrel}}{{Silva}
  et~al.}{2011}]{Silva10}
{Silva} L.,  {Schurer} A.,  {Granato} G.~L.,  {Almeida} C.,  {Baugh} C.~M.,
  {Frenk} C.~S.,  {Lacey} C.~G.,  {Paoletti} L.,  {Petrella} A.,
  {Selvestrel} D.,  2011, \mnras, 410, 2043

\bibitem[\protect\citeauthoryear{{Somerville}, {Hopkins}, {Cox}, {Robertson} \&
  {Hernquist}}{{Somerville} et~al.}{2008}]{Somerville08}
{Somerville} R.~S.,  {Hopkins} P.~F.,  {Cox} T.~J.,  {Robertson} B.~E.,
  {Hernquist} L.,  2008, \mnras, 391, 481

\bibitem[\protect\citeauthoryear{{Tuffs}, {Popescu}, {V{\"o}lk}, {Kylafis} \&
  {Dopita}}{{Tuffs} et~al.}{2004}]{Tuffs04}
{Tuffs} R.~J.,  {Popescu} C.~C.,  {V{\"o}lk} H.~J.,  {Kylafis} N.~D.,
  {Dopita} M.~A.,  2004, \aap, 419, 821

\bibitem[\protect\citeauthoryear{{Vega}, {Silva}, {Panuzzo}, {Bressan},
  {Granato} \& {Chavez}}{{Vega} et~al.}{2005}]{Vega05}
{Vega} O.,  {Silva} L.,  {Panuzzo} P.,  {Bressan} A.,  {Granato} G.~L.,
  {Chavez} M.,  2005, \mnras, 364, 1286

\bibitem[\protect\citeauthoryear{{Viola}, {Monaco}, {Borgani}, {Murante} \&
  {Tornatore}}{{Viola} et~al.}{2008}]{Viola08}
{Viola} M.,  {Monaco} P.,  {Borgani} S.,  {Murante} G.,    {Tornatore} L.,
  2008, \mnras, 383, 777

\end{thebibliography}
